\documentclass[]{article}
\pdfoutput=1
\usepackage[utf8]{inputenc}
\usepackage{jheppub}
\usepackage{graphicx}
\usepackage{amssymb}
\usepackage{amsmath}
\usepackage{bm}
\usepackage{color}
\usepackage{enumitem}
\usepackage{tensor}
\usepackage{cleveref}
\usepackage{verbatim}
\usepackage{xspace}

\DeclareMathAlphabet\mathbfcal{OMS}{cmsy}{b}{n}
\definecolor{darkgreen}{RGB}{50,150,0}
\definecolor{purple}{cmyk}{0.5,0.75,0,0}
\definecolor{darkpurple}{RGB}{128,0,128}

\newcommand{\Mpl}{M_{\rm Pl}}
\newcommand{\LCDM}{$\Lambda$CDM\xspace}

\definecolor{ultramarine}{rgb}{0.07, 0.04, 0.56}
\definecolor{cadmiumgreen}{rgb}{0.0, 0.42, 0.24}
\definecolor{indigo(dye)}{rgb}{0.0, 0.25, 0.42}
\hypersetup{
colorlinks=true,
citecolor=ultramarine,
linkcolor=cadmiumgreen,
urlcolor=indigo(dye),
pdfauthor={},
pdftitle={},
pdfsubject={}
}

\title{Rock `n' Roll Solutions to the Hubble Tension}

\author{Prateek Agrawal$^{1,2}$,}
\author{Francis-Yan Cyr-Racine$^{2,3}$,}
\author{David Pinner$^{2,4}$, and}
\author{Lisa Randall$^{2}$}
\affiliation{$^{1}$Rudolf Peierls Centre for Theoretical Physics, University of Oxford, Parks Road, Oxford OX1 3PU, United
Kingdom}
\affiliation{$^{2}$Department of Physics, Harvard University, 
17 Oxford St., Cambridge, MA 02138, USA}
\affiliation{$^{3}$Department of Physics and Astronomy, University of New Mexico, 
210 Yale Blvd NE, Albuquerque, NM 87106, USA}
\affiliation{$^{4}$Department of Physics, Brown University, 
182 Hope St., Providence, RI 02912, USA}

\begin{document}
\abstract{ Local measurements of the Hubble parameter are increasingly
  in tension with the value inferred from a $\Lambda$CDM fit to the
  cosmic microwave background (CMB) data.  In this paper, we construct
  scenarios in which evolving scalar fields significantly ease this
  tension by adding energy to the Universe around recombination in a
  narrow redshift window. We identify solutions with scalar field potential $V \propto \phi^{2
  n}$ that have simple asymptotic behavior, both oscillatory (rocking) and
  rolling. These solutions consistently describe both the 
  field evolution and its fluctuations without approximation. Our findings differ qualitatively from
  some of the existing literature, which rely upon a coarse-grained
fluid description.  Combining CMB data with low-redshift
  measurements, the best fit model has $n=2$ with a significantly higher value of the Hubble constant as compared to a $\Lambda$CDM fit to the same data. Future measurements of the late-time amplitude of
  matter fluctuations and of the reionization history could help
  distinguish these models from competing solutions.  
} 

\emailAdd{prateek.agrawal@physics.ox.ac.uk}
\emailAdd{fycr@unm.edu}
\emailAdd{randall@physics.harvard.edu}

\maketitle

\section{Introduction} \label{sec:intro}

The concordance model, $\Lambda$CDM, has been
very successful in describing the entire observed
history of the Universe. But we are now entering the era of precision cosmology where
the basic picture can be placed under closer scrutiny. Notably, as the
precision of measurements has improved, hints of cracks in the
concordance model have appeared.  The discrepancy of the present day
Hubble parameter $H_0$ derived from local measurements
\cite{riess11,freedman2012,Riess:2016jrr,Suyu:2016qxx,Riess:2018byc,Bonvin:2016crt,Dhawan:2017ywl,2019MNRAS.484.4726B,Soares-Santos:2019irc,Riess:2019cxk,Wong:2019kwg,Huang:2019yhh,Kourkchi:2020iyz, Reid:2019tiq,freedman2019,Freedman:2020dne,Pesce:2020xfe, Khetan:2020hmh, Blakeslee:2021rqi,Birrer:2020tax,Riess:2021jrx,Freedman:2021ahq} and from cosmic microwave background (CMB)
measurements \cite{Hinshaw:2012aka,planck15,Aghanim:2018eyx} 
is perhaps the most notable indication of a problem. This  long-standing deviation (see e.g.~refs.~\cite{Freedman:2000cf,Verde:2019ivm,DiValentino:2021izs,Perivolaropoulos:2021jda,Abdalla:2022yfr}) has become stronger with
time and is currently
at a very high statistical significance level \cite{Riess:2021jrx}.  
This is worth taking seriously,
both for itself and as a template for how we will view current and
future cosmological measurements and their potential to probe new
physics.
While
the disagreement may ultimately be explained by experimental
systematics (see
e.g.~refs.~\cite{Rigault:2013gux,Rigault:2018ffm,Jones:2018vbn,Shanks:2018rka,Riess:2018kzi,Kenworthy:2019qwq,Efstathiou:2020wxn,Mortsell:2021nzg,Mortsell:2021tcx})
or a statistical fluctuation, it is possible that we are
seeing the first signs of a theory beyond the standard cosmological
model.

Through the positions and spacing of the acoustic peaks present in CMB  power spectra, the CMB is sensitive to $H_0$ via the
well-measured angle $\theta_{\rm s} = r_{\rm s} / D_{\rm M}(z_*)$, where
$r_{\rm s}$ is the
comoving sound horizon at decoupling, and $D_{\rm M}(z_*)$ is the
comoving angular
diameter distance to the surface of last scattering. The sound horizon
is sensitive to early physics, whereas $D_{\rm M}$ is sensitive to physics
after photon-baryon decoupling,
\begin{align}
  r_{\rm s}
  &=
  \int_{z_*}^\infty
  \frac{c_{\rm s} dz'}{ H(z')},
  \\
  D_{\rm M}(z)
  &=
  \int_{0}^{z}
  \frac{dz'}{ H(z')}.
  \label{eq:rsDA}
\end{align}

Baryon acoustic oscillation (BAO) data at lower redshifts provide another measurement of 
$r_{\rm s} / D_{\rm M}$, which asymptotes to $r_{\rm s} H_0/z$ for small $z$. 
This implies that in order to accommodate a value of $H_0$ that is
compatible with the
local measurements (and larger
than the Planck \LCDM best fit \cite{Aghanim:2018eyx}), $r_{\rm s}$ has to be
modified \cite{Evslin:2017qdn,2019ApJ...874....4A,knox20}.

One simple way to reduce $r_{\rm s}$ is to increase $H(z)$ before
decoupling\footnote{One could also reduce the sound horizon by decreasing the sound speed of the cosmic plasma, or by having hydrogen recombination occur earlier \cite{Jedamzik:2020krr,Chiang:2018xpn,Hart:2019dxi,Sekiguchi:2020teg}.}. Since the integral for $r_{\rm s}$ is dominated by contributions
just before the epoch of CMB last-scattering, energy injection around that time is the most
efficient in reducing $r_{\rm s}$.
Indeed, energy injection that is peaked in a narrow window of redshift
around recombination minimizes
the impact on other successful \LCDM
predictions~\cite{Hojjati:2013oya,Poulin:2018dzj,
2019ApJ...874....4A,knox20}. Along those lines, there have been a number of studies that try to resolve the Hubble
tension by extending \LCDM by a new energy contribution with
distinctive time-dependence and properties (see, e.g.,~refs.~\cite{DiValentino:2017rcr,DEramo:2018vss,Poulin:2018zxs,Pandey:2019plg,Mortsell:2018mfj,Vattis:2019efj,Renk:2017rzu,Khosravi:2017hfi,Yang:2018uae,Yang:2018euj,Yang:2018qmz,Kreisch:2019yzn,Benevento:2020fev}). One such scenario, often referred to in the literature as ``early dark energy'' \cite{Hill:2018lfx,Karkare:2018sar,Poulin:2018cxd,NobleChamings:2019ody,Smith:2019ihp,Bhattacharyya:2019lvg,Lin:2019qug,Lin:2020jcb,Yin:2020dwl,Chudaykin:2020acu,Ivanov:2020ril,Khoraminezhad:2020cer,Hill:2020osr,Sakstein:2019fmf,Mandal:2019adw,Niedermann:2020dwg,Ye:2020oix,Braglia:2020bym,Garcia:2020sjl,Freese:2021rjq,Klypin:2020tud,Seto:2021xua,Weiner:2020sxn,Tian:2021omz,Gomez-Valent:2021cbe,Smith:2020rxx,Murgia:2020ryi,Gogoi:2020qif,Chudaykin:2020igl,Nojiri:2021dze,CarrilloGonzalez:2020oac,Niedermann:2019olb,Niedermann:2020qbw,Jiang:2021bab,Ye:2021iwa,Hill:2021yec,Karwal:2021vpk,McDonough:2021pdg,LaPosta:2021pgm,Gomez-Valent:2022bku,Kojima:2022fgo,Oikonomou:2022yle,Smith:2022hwi,Niedermann:2021vgd,Niedermann:2021ijp,Chang:2021yog,Sabla:2022xzj,Herold:2021ksg,Wang:2022jpo,Jiang:2022uyg,MohseniSadjadi:2022pfz,Reeves:2022aoi,McDonough:2022pku,Simon:2022adh,Wang:2022nap,Cruz:2022oqk,Trodden:2022zye,Nakagawa:2022knn,Murai:2022zur,Hart:2022agu,Hayashi:2022eha,Herold:2022iib,Takahashi:2022cpn,Lin:2022phm,Haridasu:2022dyp,Rudelius:2022gyu,Maziashvili:2021mbm,Brissenden:2023yko}, involves a light scalar field whose dynamics can increase $H(z)$ prior to recombination (see ref.~\cite{Kamionkowski:2022pkx} for a review).

Evolving scalar fields feature prominently in models of inflation and quintessence (see e.g.~refs.~\cite{Ferreira:1997hj,Tsujikawa:2013fta}). 
In this paper we present a class of scalar
field potentials and solutions which injects energy
in the requisite redshift window close to recombination to shrink the sound horizon. 
We present simple, but exact, solutions that are
sufficient to provide the energy injection needed and are under
enough control that we can trace their evolution robustly. These
solutions arise from potentials of the form $V\propto \phi^{2n}$, with
$n$ determining whether the solutions are asymptotically oscillatory
or not. In particular, we identify solutions with asymptotically constant equation of state for large $n$, where there are no oscillatory
solutions. 

In contrast to some previous analyses (see, e.g.,refs.~\cite{Karwal:2016vyq,Poulin:2018dzj,Poulin:2018cxd}), we use the full equations of
motion for the background as well as for the fluctuations, which allows us to show that the coarse-grained approximation used in some past works
is not valid near the peak of the energy injection. We will
show that these models can fit the CMB and the SH0ES measurement
better than \LCDM, with the power-law index $n=2$ mildly preferred.

The outline of the paper is as follows.  In
section~\ref{sec:scaling_solns} we introduce the class of scalar
potentials which produce brief energy injections in the early
universe.  We classify the solutions to the scalar field equations of
motion and discuss their stability.  In
section~\ref{sec:cosmo_results} we describe the procedure we use to
fit the models to cosmological data and present the results, comparing
to alternate solutions.  We highlight unresolved issues in
section~\ref{sec:disc}, commenting on their potential resolution,
before concluding in section~\ref{sec:conc}.

\section{Classification of scalar field solutions in cosmology}\label{sec:scaling_solns}

CMB data require that 
the ratio of energy density of the scalar
field to the background energy density should peak in a
relatively small window in 
redshift~\cite{Hojjati:2013oya,Poulin:2018dzj,Poulin:2018cxd}.
The goal of this section is to demonstrate
potentials that give rise to this behavior.

The energy-momentum tensors for uncoupled fluids are separately covariantly conserved. 
A consequence of this conservation law is that the ratio of energy
densities is determined by the relative equation of
state parameters of the scalar field ($w_\phi$) and the background
($w_b$):
\begin{align}
  \frac{\rho_{\phi}}{\rho_{b}} 
  &\propto 
  \exp\left({-\int 3[w_{\phi}(a) - w_b(a)] d \ln a}\right)
  \,.
\end{align}
In order that the energy density in the scalar field does not come to
dominate that of the background either at early or late
times, $(w_{\phi} - w_b)$ must transition from
negative to positive. 
This energy injection, which is required to be $\mathcal{O}(10\%)$ in order to resolve the Hubble tension, is
localized at the redshift when $w_\phi = w_b$.

One particularly simple class of such solutions has the scalar field
initially frozen on its potential, with $w_{\phi} \approx -1$, before
thawing onto a trajectory with a nearly constant equation of
state, $w_{\phi} > w_b$, which we will refer to as rolling solutions.  This behavior is in contrast with the
familiar case of, {\it e.g.}, axion dark matter, in which the field
and its equation of state oscillate rapidly after thawing.
The axion
oscillation frequency grows relative to the decreasing Hubble
parameter, so that it can become intractable to trace the oscillating
field evolution
numerically. 

Cycle-averaged coarse-graining \cite{Turner:1983he} can provide an accurate description of
the fluid at late times, both in the case of axion dark matter as well
as more general potentials with oscillating equation of state.
However, the coarse-graining approximation does not apply
near the initial thawing phase when the field starts oscillating, and
this is precisely the regime in which the details of the
field evolution and fluctuations have the greatest impact on the
gravitational potential, and thence the CMB. Therefore, a
cycle-averaged description for such an oscillating solution might be
insufficient.

This is why we look for solutions that quickly asymptote to a constant
$w_{\phi}$. These have been studied in a different cosmological context in refs.~\cite{Ratra:1987rm,Liddle:1998xm,Dutta:2008px}.
The rolling trajectory is easiest to describe in a simplified limit,
in which the background cosmology has a constant equation
of state $w_b$. Further,
we work in the approximation where we neglect the back-reaction of the
scalar field.
This should be justified since the maximal energy injection
that we are interested in will be $\mathcal{O}(10\%)$. 
We emphasize that we use these simplifications to elucidate the
physics of the scalar field and do not use them for any of our
numerical results.

\subsection{Emden-Fowler solutions}\label{subsec:E-F}
We look for a potential such that the late time solution 
is
a trajectory where
the scalar field rolls with a constant equation of state, $w_{\phi}>w_b$,
\begin{align}
  \rho_\phi(a)
  &=
  \rho_0 
  \left(\frac{a_0}{a}\right)^{3(1+w_{\phi})}.
  \label{eq:rhophi}
\end{align}
Recalling that $1+w_\phi = a^2 H^2 (\partial_a \phi)^2 / \rho_\phi$, 
we can extract
formulae for $V(\phi)$ and $\partial_a{\phi}$
\begin{align}
  V(\phi) &= \frac{1-w_{\phi}}{2} \rho_\phi,
  \\
  \partial_a \phi &= \frac{\sqrt{(1+w_{\phi})\rho_\phi}}{a H}.
\end{align}
Using $3 H^2 \Mpl^2 \approx \rho_b = \rho_{b0}
\left(\frac{a_0}{a}\right)^{3(1+w_b)}$, we can solve for $\phi(a)$,
\begin{align}
  \phi(a)
  &=
  c \left(\frac{a_0}{a}\right)^{\frac32 (w_\phi-w_b)},
  \qquad
  c = \frac{\Mpl}{(w_\phi-w_b)}\sqrt{\frac{4(1+w_\phi)\rho_0}{3\rho_{b0}}},
\end{align}
where we have chosen the additive constant for the solution to correspond to
$\phi=0$ at $a=\infty$.
Since both the potential as well as the field have power-law
dependence on $a$, we see that the potential is a power-law in $\phi$ as
well. Explicitly,
\begin{align}
  V(\phi)
  &=
  \frac12(1-w_\phi) \rho_0 \left(\frac{\phi}{c}\right)^{2n},
  \quad n =
  \frac{1+w_\phi}{w_\phi-w_b}.
\end{align}
These $\phi^{2n}$ potentials have been previously studied in the
context of axion models \cite{Karwal:2016vyq,Poulin:2018dzj} and quintessence \cite{Liddle:1998xm,Dutta:2008px}, but we emphasize that we
have arrived at them independently from theoretical motivation by
demanding the rolling solutions.
\begin{figure}[t]
  \centering
  \includegraphics[width=0.69\textwidth]{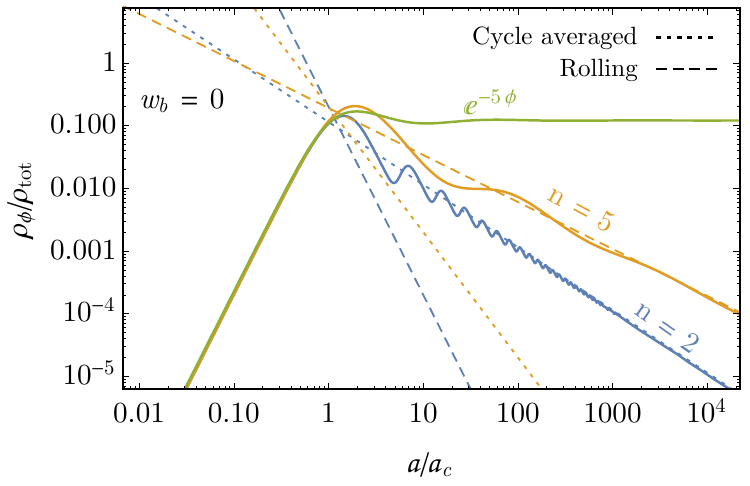}\\
  \includegraphics[width=0.69\textwidth]{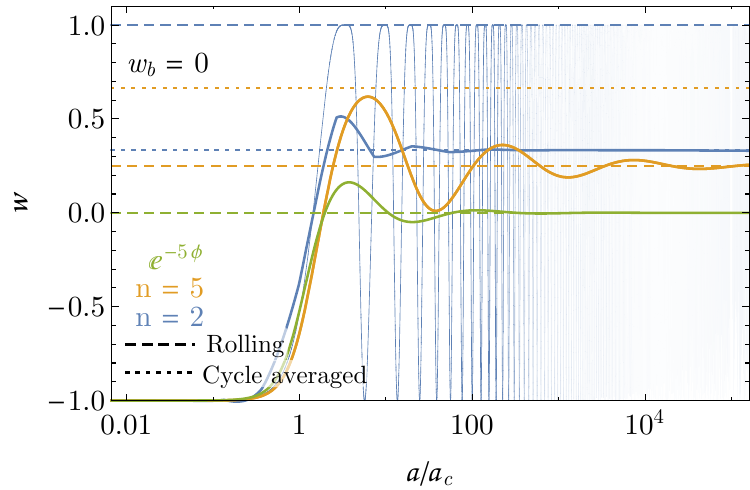}
  \caption{
    \emph{Top panel:} Evolution of the energy density of the scalar field $\phi$
    relative to total energy density of the universe in a matter-dominated toy
    cosmology ($w_b=0$). Here, $\rho_{\rm tot}$ includes the contribution from the scalar field, and $a_c$ denotes the scalar factor at which the field starts rolling. We show the
    evolution for different choices of $n$ in the scalar potential
    $V\propto \phi^{2n}$, as well as an example of exponential potential,  $V\propto e^{-\lambda\phi}$. 
    We choose values of $n$ to show an example each of an oscillating ($n=2$)
    and a rolling solution ($n=5$).  The asymptotics of the numerical solutions is
    shown to match well with the analytical results derived in the
    text, neglecting the back-reaction.  Note that the presence of slow oscillations around the rolling solution for $n=5$ is due to the transition from $w_{\phi} = -1$ to the asymptotic solution; the stability of the rolling solution guarantees that these oscillations are quickly damped. We note that the width of the energy injection grows larger as $n$ is increased. \emph{Bottom panel:} Equation of state of the scalar field as a function of the scale factor for the same potentials as in the left panel. For $n=2$, the cycle-average of $w_\phi$ (dark thick blue line) rapidly approaches $w_{\rm osc}=1/3$, while for $n=5$, it asymptotes to the rolling solution with $w_{\phi}=1/4$ instead of the cycle-averaged approximation ($w_{\rm osc}=2/3$). For the exponential potential, the scalar field approaches the well-known tracking behavior with $w_\phi=w_b$. 
  }
  \label{fig:emden-fowler}
\end{figure}

Even when these rolling solutions exist, they may not be the unique
solutions. Fortunately,
the equation of motion for the $\phi^{2n}$ potential can be cast into
Emden-Fowler form, the solutions of which have been studied in detail
(see~\cite{10.2307/2029180} for a review). 
The equation of motion of $\phi$ on a constant $w_b$ background, with
back-reaction neglected, is
\begin{align}
  \frac{\partial^2 \phi}{\partial(\log a)^2}
  + \frac32 (1-w_b) \frac{\partial \phi}{\partial (\log a)}
  + \frac{\partial_\phi V}{H^2(a)}
  =0.
\end{align}
Identifying $s \propto a^{\frac32(1-w_b)} $ and $y= s \phi$, the
above equation reduces to the Emden-Fowler form, 
\begin{align}
  y''(s) + s^\sigma y^\gamma(s) &= 0,
\end{align}
with $\sigma =
\frac{4}{1-w_b}-2(n+1)$
and $\gamma = 2n-1$. The solutions of Emden-Fowler are characterized
by the values of $\sigma$ and $\gamma$. The different regimes of the solution are listed in table \ref{tab:Enden_Fowler_solns}. 

\begin{table}
\begin{align}
  \begin{array}{|c|c|c|c|c|}
    \hline
    \text{Asymptotic solutions}& \text{Emdem-Fowler} 
    &\text{Translation to} &
    \multicolumn{2}{c|}{\text{Background}}
    \\
    \cline{4-5}
    &\text{conditions} &\text{scalar field models}&
    \text{Radiation}
    &
    \text{Matter}
    \\
    \hline
    \text{Osc. only}
&
\sigma + 2 \geq 0
& 
n \leq \frac{2}{1-w_b}
& n \leq 3 & n \leq 2\nonumber
\\
\vspace{-3mm}
&&&&\\
\text{Osc. 
+ non-osc.}
&
\sigma +2 < 0 \leq \sigma + \frac{\gamma+3}{2}
&
\frac{3+w_b}{1-w_b} \geq n > \frac{2}{1-w_b}
& 5\geq n>3& 3\geq n>2
\\
\vspace{-3mm}
&&&&\\
    \text{Non-osc. only}
    &\sigma + \frac{\gamma+3}{2} < 0
& n > \frac{3+w_b}{1-w_b}
& n > 5
& n > 3
\\
\hline
\end{array}
\end{align}
\caption{Asymptotics of solutions to the Emden-Fowler equation, and
their translation to the parameterization used in this paper. The last two columns give the condition for a given type of solution to exist in either radiation- and matter-dominated cosmologies, respectively. \label{tab:Enden_Fowler_solns}}
\end{table}

The existence of these rolling trajectories is associated with the fact that $\partial^2_\phi V(\phi) /H(a)^2$ is an
$\mathcal{O}(1)$ constant on these solutions,
\begin{align}
  \frac{\partial_\phi^2 V(\phi)}{H(a)^2}
  &=
  \frac94 (1-w_\phi)(2+w_\phi+w_b)
  \,,
\end{align}
so that the curvature of the potential never grows large relative to the expansion rate of the universe.  This should be contrasted with the oscillatory case, \emph{e.g.}, $n=1$.  In this case $\partial_{\phi}^2 V(\phi)$ is initially much smaller than the expansion rate, and the field is correspondingly stuck on its potential.  As the expansion rate decreases, the field begins to roll when $\partial_{\phi}^2 V(\phi)$ and $H(a)^2$ become comparable, eventually oscillating rapidly at a frequency controlled by $\partial_{\phi}^2 V(\phi) \gg H(a)^2$.  The fact that $\partial_{\phi}^2 V(\phi)$ never becomes large compared to $H(a)^2$ on the rolling solutions thus prevents the onset of oscillations.

The lower bound on $n$ for non-oscillating solutions to exist agrees
with the restriction that $w_\phi \leq 1$, given that 
$n = (1+w_{\phi})/(w_{\phi}-w_b)$ on the rolling solution.  As $n$ is
increased above this boundary, both oscillatory and rolling solutions
exist simultaneously.  However, they are not equally relevant:
as we will show in section~\ref{ssec:stability}, the rolling solutions
are unstable whenever both oscillatory and rolling solutions exist.
Rapidly oscillating solutions can be 
modeled by a cycle-averaged description with the effective equation of state given by~\cite{Turner:1983he,Poulin:2018dzj}
\begin{align}
  w_{\rm osc}
  \simeq \frac{n-1}{n+1}.
\end{align}

While tracking these oscillatory solutions 
could in principle prove problematic for numerical
evaluation, in practice these solutions never undergo more than
$\mathcal{O}(10)$s of cycles before the oscillations are cut off.
This is because of the transition from matter to dark energy
domination in the late Universe.  As demonstrated by the boundaries
given in table~\ref{tab:Enden_Fowler_solns}, when the background
equation of state approaches $w_b \rightarrow -1$, all solutions are
non-oscillatory for $n > 1$.  As a result, we are able to exactly
track these solutions in our numerical results in the following
sections, without relying on cycle-averaged approximations.

\subsection{Exponential potentials}

Exponential potentials are a special limiting case of the rolling solutions which occur when $w_\phi = w_b$, corresponding to the limit $n\rightarrow\infty$. In this
case, the field $\phi$ depends
logarithmically on $a$,
\begin{align}
  \phi(a)
  &= \phi_0 + \sqrt{\frac{3(1+w_b)\rho_0}{\rho_{b0}+\rho_0}}\log\frac{a}{a_0},
  \qquad\qquad(w_\phi=w_b).
\end{align}
In this case we have kept the back-reaction of the field since 
it is possible to obtain a simple analytical solution even with the
back-reaction included.
This trajectory corresponds to an exponential potential, 
\begin{align}
  V(\phi)
  &=
  V_0 \exp \left(-\lambda \frac{\phi}{\Mpl}\right),
  \quad
  \lambda = 
  \sqrt{
    3(1+w_b)
    \left(
      1+\frac{\rho_{b0}}{\rho_{0}}
    \right)
  },
  \,
  \qquad (w_\phi=w_b).
  \label{eq:exp-pot}
\end{align}
On this solution,
\begin{align}
  \frac{\partial_\phi^2 V(\phi)}{H(a)^2}
  &=
  \frac92(1-w_b^2),
  \qquad\qquad
  (w_\phi=w_b)
\end{align}
is an $\mathcal{O}(1)$ constant, as for the non-oscillatory solutions discussed in the previous subsection. These exponential potentials have previously been studied in the context of quintessence models~(see~\cite{Tsujikawa:2013fta} for a review).  However, since $w_\phi=w_b$, the energy injection for this potential does
not redshift relative to the background~(see figure~\ref{fig:emden-fowler}), which prevents these solutions from being ideal candidates to resolve the Hubble tension.  Therefore we will focus on the case of monomial potentials at finite $n$ for the remainder of the paper.

\subsection{Stability of fluctuations}
\label{ssec:stability}

As demonstrated in the previous subsection, for $\phi^{2n}$ potentials in the range $(3 + w_b)~\geq~n~(1 - w_b)~>~2$, rolling solutions are irrelevant at late times, since the field redshifts more quickly than the oscillation amplitude.  In fact, the rolling solutions are unstable for this range of $n$.  The homogeneous equation of motion for linear fluctuations about the rolling solution, written using $\log{a}$ as the time variable and assuming a background equation of state $w_b$, is
\begin{align}
  \frac{\partial^2 \delta \phi_k}{\partial(\log a)^2}
  + \frac32 (1-w_b) \frac{\partial \delta\phi_k}{\partial (\log a)}
  + \left[\frac{k^2}{a^2 H(a)^2} + \frac{9}{4} (1 - w_{\phi}) (2 + w_{\phi} + w_b) \right] \delta \phi_k
  =0.
\end{align}
It has solutions which scale as
\begin{align}
\delta \phi_0 \sim a^{-\frac{3}{4}\left[(1 - w_b) \pm \sqrt{4 w_{\phi}^2 + (1 + w_b) (4 w_{\phi} + w_b - 7)}\right]}.
\end{align}
For $w_{\phi} < \sqrt{2} \sqrt{1+w_b} - (1+w_b)/2$, these solutions are oscillatory, and their envelope redshifts as $a^{-3(1-w_b)/4}$, while the rolling solution redshifts as $a^{-3(w_{\phi} - w_b)/2}$.  Thus the fluctuations grow relative to the rolling solution for $1 > w_{\phi} > (1 + w_b)/2$, corresponding to $2 < n~(1 - w_b) < 3 + w_b$ and coinciding with the entire range of $n$ for which both oscillatory and rolling solutions exist.  For larger $n$, these fluctuations redshift more quickly, and the rolling solutions are stable.

The oscillatory solutions, on the other hand, exhibit a somewhat more subtle instability.  For $n > 1$, $\partial^2_{\phi} V \propto \phi^{2 (n-1)}$ vanishes as $\phi \rightarrow 0$.  During each cycle of the oscillation, the fluctuations of the $\phi$ field become massless, leading to a short burst of particle production.  If the growth rate of the instability is sufficiently large, $\partial^2_{\phi} V \gg H^2$, the field will eventually fragment completely into modes with $k > 0$~\cite{Lozanov:2017hjm,Johnson:2008se}, invalidating the linear perturbation theory used in our analysis.

On the oscillating solutions,
\begin{align}
  \left.  \frac{\partial_{\phi}^2 V}{H^2}\right|_{osc}
  &\propto
  a^{-\frac{3}{n+1}\left(n(1 - w_b) - 3 - w_b\right)},
\end{align}
so that the growth rate increases with scale factor for $n < (3 + w_b)/(1 - w_b)$.  The case of $w_b = 0$ is most relevant for our numerical results, since the shape of the energy injection constrains most of the oscillations to occur during matter domination.  For $n=3$, the growth rate of the fluctuations remains smaller than the expansion rate at all times, so that they never become important.  For $n=2$, however, the growth rate accelerates, and it is a numerical question whether these fluctuations become important before the onset of dark energy domination, at which point the background solutions cease to be oscillatory.  Of course, a realistic potential may include a mass term for the scalar field which could prevent this instability entirely\footnote{From a model-building perspective, such a mass term would need to be stabilized against  UV corrections by, for example, a shift-symmetry.}.  For our results in the following sections, we have not included such a mass term; rather, we have simply checked numerically that the scalar field fluctuations remain under control.

\subsection{Comparison with coarse-grained description}

The oscillatory solutions have a frequency which monotonically increases relative to the Hubble expansion rate, which makes it numerically challenging to track their evolution to late times. The usual method for addressing this issue is to coarse-grain these solutions, along with their fluctuations, replacing the numerically computed equation of state  and rest-frame sound speed by their cycle-averaged values~\cite{Turner:1983he,Poulin:2018dzj},
\begin{align}
w_{\rm osc} \simeq \frac{n-1}{n+1}, &\hspace{1 cm} \langle c_s^2 \rangle \simeq \frac{2 a^2 (n - 1) \omega^2 + k^2}{2 a^2 (n+1) \omega^2 + k^2},
\end{align}
in which $\omega$ is the instantaneous oscillation frequency of the field fluctuations.  As demonstrated in figure~\ref{fig:emden-fowler}, however, this description is inaccurate near the onset of thawing, when the ratio of energy density of the scalar field to that of the background is near its maximum.  In fact, we will see (figure \ref{fig:energy_injection} below) that the oscillating solutions go through a short period of kination ($w \approx 1$) during the first oscillation cycle, rapidly depleting their energy density relative to the cycle-averaged solution.  For $n=2$ and $w_b = 0$, it takes several oscillation cycles before the oscillation frequency becomes large relative to the Hubble rate, at which point $\rho_{\phi} / \rho_{\text{tot}}$ has dropped by more than two orders of magnitude relative to its $\mathcal{O}(10\%)$ maximum.  

The fluctuations are similarly poorly described by coarse-graining.  At the onset of thawing, $\partial_{\phi}^2 V \simeq H^2$, so that all subhorizon modes have $k^2 \gtrsim \partial_{\phi}^2 V$.  Thus the fluctuations are kinetic energy dominated, and their rest-frame sound speed is close to 1 for the entire relevant range of scale factors and $k$-modes.  By contrast, taking the above cycle-averaged formula gives $\langle c_s^2 \rangle \rightarrow w_{\rm osc} < 1$ over a large range of $k$, since $\omega$ becomes large relative to $H$ at late times.  The detailed evolution of these fluctuations qualitatively affects the fit of these scalar field models to the CMB data; therefore, in our numerical study, we have chosen not to make any coarse-graining approximations.  For the following results, we have tracked the individual oscillations to late times, checking that numerical errors remain under control until the energy density of the scalar field is sufficiently subdominant.

\section{Cosmological Analysis}\label{sec:cosmo_results}
\subsection{Parameterization}
In this section we present the fits for the models above to the
cosmological
data. We use the following parametrization for the model,
\begin{align}\label{eq:potential}
  V(\phi)
  &=
  V_0 \left(\frac{\phi}{\Mpl}\right)^{2n} + V_{\Lambda},
\end{align}
where $V_{\Lambda}$ is a constant. In addition to $V_0$, $V_{\Lambda}$, and $n$, the scalar field initial conditions $\phi_i$ and $\partial_a\phi_i$ are
the two other
free parameters. Since we are primarily interested in field solutions that are initially frozen, we will always set the velocity of the scalar field
to zero at very early times.  Given $\phi_i \sim M_{\rm pl}$ initial
conditions, the field will then begin to thaw when $V_0 \sim \rho_b$.
Thus in order that the field begins to roll near matter-radiation
equality, we must have $V_0 \sim {\rm eV}^4$.  Note that the smallness
of the dimensionless coupling, $({\rm eV}/M_{\rm pl})^4$, is
consistent with a softly broken shift symmetry (from e.g.~instantons) and therefore
technically natural, similar to the case of the $\Lambda_{\rm QCD}^4 \cos{(a/f_a)}$ potential for the QCD axion.  For a discussion of the tuning required to
explain the absence of additional couplings, see
section~\ref{sec:disc}.  In our model, the value of $V_{\Lambda}$
determines the late-time dark energy density, and its value is set by
demanding that the Universe is spatially flat. The parameter $n$
determines the shape of
the energy injection.  We note that eq.~\eqref{eq:potential}
completely determines the evolution of the scalar field and its
perturbations. In particular, there is no extra freedom to
independently change the sound speed of the scalar field perturbations
in this model. 

In principle, we could  perform our Markov Chain Monte Carlo (MCMC) analysis in terms  of $V_0$ and $\phi_i$. However, since our preliminary analysis shows that cosmological data are primarily sensitive to both the timing and the amount of energy injected by the rolling scalar field, we choose to parametrize the model using the approximate redshift at which the field starts rolling, $z_{\rm c, MC}$, and the approximate fraction of energy injected by the scalar field $f_{\phi,{\rm MC}}\approx (\rho_\phi / \rho_{\rm tot})|_{\rm max}$. 
The initial field value $\phi_i$ and the $V_0$ parameter are then
determined through the relations $\partial^2_\phi V(\phi_i) \equiv
9 H^2(z_{\rm c, MC})$ and $V(\phi_i)\equiv 3 f_{\phi,{\rm MC}}
H^2(z_{\rm c, MC}) \Mpl^2$. These latter relations should be interpreted as the definitions of $z_{\rm c, MC}$ and $f_{\phi,{\rm MC}}$.

We implement this model into a modified version of the \texttt{Class} code \cite{2011arXiv1104.2932L,2011JCAP...07..034B}. We perform parameter scans over $f_{\phi,{\rm MC}}$ and $\ln{(1+z_{\rm c, MC})}$ as well as the standard six flat $\Lambda$CDM parameters using the \texttt{MontePython} \cite{Brinckmann:2018cvx,Audren:2012wb} software package version 3.1\footnote{This was the most-recent version available at the onset of the work.}.  We use flat prior distributions on our scalar field parameters, $f_{\phi,{\rm MC}}\in [10^{-4},0.3]$\footnote{We set a prior on $f_{\phi,{\rm MC}}$ rather than its logarithm to avoid volume effects at low $f_{\phi,{\rm MC}}$ values (see e.g.~refs.~\cite{Smith:2020rxx,Herold:2021ksg,Herold:2022iib,Gomez-Valent:2022hkb,Reeves:2022aoi}). The lower bound is chosen to avoid numerical instabilities arising when $f_{\phi,{\rm MC}}=0$ while allowing the model to essentially reduce to $\Lambda$CDM in this limit.} and $\ln{(1+z_{\rm c, MC})}\in [7.5,9.5]$, and the same prior distributions for the other cosmological parameters as in the Planck analysis \cite{planck15}. We solve both the background and perturbed Klein-Gordon equations at each step to obtain the exact cosmological evolution of the scalar field and its perturbations, and use this solution to determine a posteriori the exact maximum fraction of energy injected by the rolling field, $f_\phi$, as well as the critical redshift $z_{\rm c}$ at which the energy injection peaks. In the following, we shall present our results in terms of these latter exact quantities. 

\subsection{Datasets} 
In our cosmological analysis, we use the following datasets\footnote{These datasets were the latest available at the time the analysis was performed.}:
\begin{itemize}
  \item {\bf Planck}:
    We use the Planck 2015 CMB temperature and polarization likelihoods \cite{Aghanim:2015xee} for both low-$\ell$ and  high-$\ell$, including all nuisance parameters and using the same prior distributiona as in the Planck analysis \cite{planck15} for those. We also include the Planck lensing likelihood \cite{plancklens}.  

\item {\bf BAO}: We include baryon acoustic oscillation (BAO) measurements from the CMASS and LOWZ galaxy samples of the Baryon Oscillation Spectroscopic Survey (BOSS) DR12 \cite{Alam:2016hwk}. These include both line-of-sight and transverse BAO measurements, as well as constraints on the growth of structure through the parameter $f\sigma_8$. We shall refer to these datasets as high-$z$ BAO. We also include low-$z$ BAO measurements from the 6dF Galaxy Survey \cite{BAO1} and from the Main Galaxy Sample (MGS) of SDSS \cite{BAO3}. 

\item {\bf SH0ES}:
We include the measurement \cite{Riess:2018byc} of the local Hubble parameter $H_0 = 73.52\pm1.62$ km/s/Mpc\footnote{As discussed in refs.~\cite{Camarena:2021jlr,Efstathiou:2021ocp}, using a Gaussian prior on $H_0$ can lead to inconsistencies when considering cosmological models that differ significantly from $\Lambda$CDM at late times. Since the models considered here primarily affect cosmology at early times, we can safely use a Gaussian prior on $H_0$.}.

\item {\bf Pantheon}: We include the Pantheon compilation of type Ia supernovae (SN Ia) \cite{Scolnic:2017caz}, which includes 1048 luminosity distances in the redshift range $0.01<z<2.3$. We include in our fit the nuisance parameter $M$ describing the fiducial absolute magnitude of a SN Ia. 

\end{itemize}

\subsection{Results: Fixed integer values of \texorpdfstring{$n$}{n}}\label{sec:res_fixed_n}

\begin{figure}[t!]
  \centering
  \includegraphics[width=\textwidth]{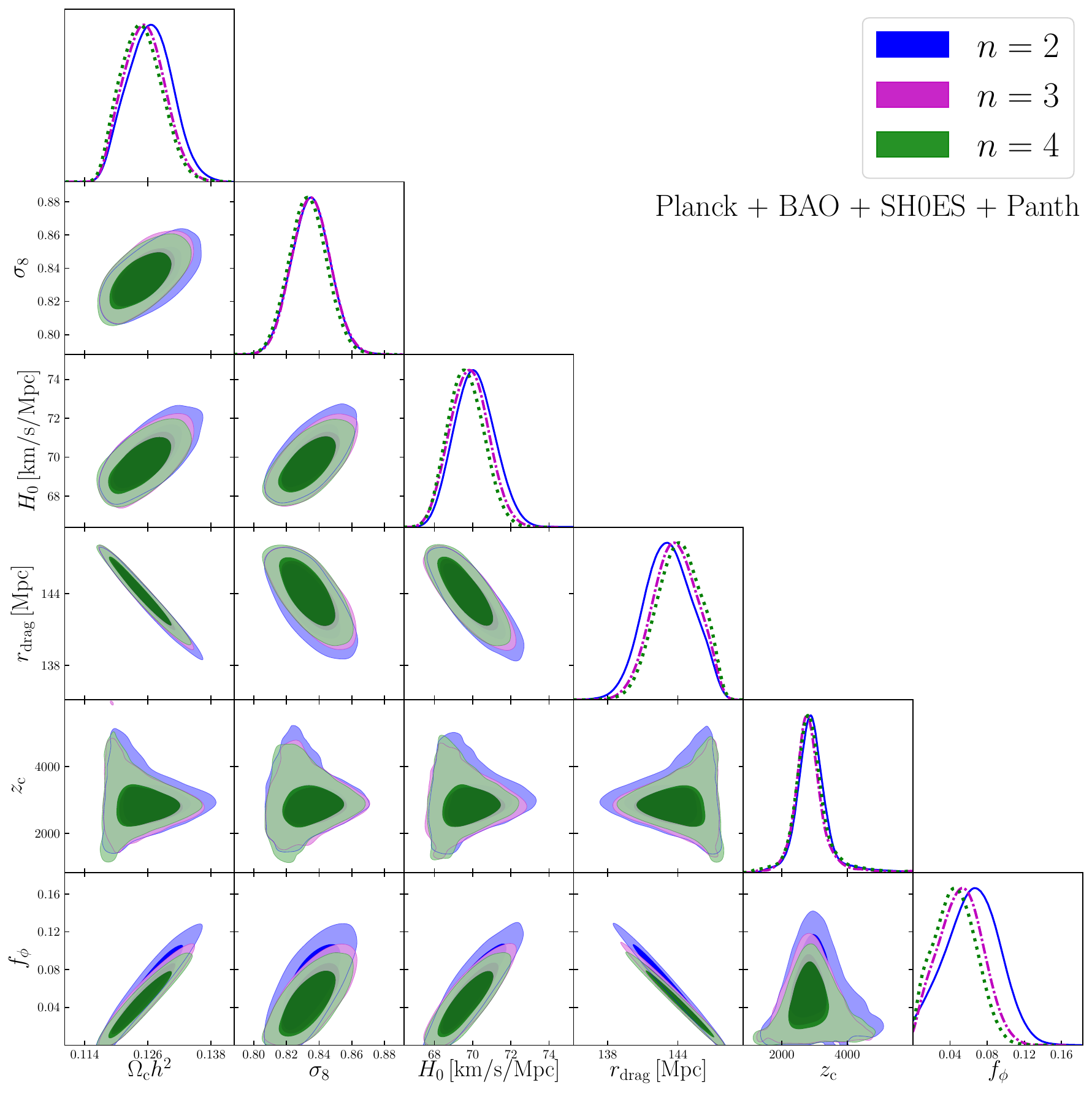}
  \caption{Marginalized posterior distributions for $V\propto\phi^{2n}$ models for three different values of $n$. Results are shown here for the data combination ``Planck + BAO + SH0ES + Pantheon''. }
  \label{fig:triangle_for_diff_n}
\end{figure}

We show posterior distributions for the most relevant parameters in figure \ref{fig:triangle_for_diff_n} for $n=2$, $3$, and $4$. The most striking feature is the similarity of the different distributions as $n$ is changed, with the notable exception of the injected energy fraction $f_\phi$. Of the models shown, the $n=2$ (corresponding to $V(\phi)\propto \phi^4$) posterior stands out for being more highly weighted towards larger values of the energy injection, which results in slightly larger values of the Hubble constant being preferred. Indeed, the inclusion of the SH0ES data in the analysis leads to a nearly 2$\,\sigma$ preference for a nonzero energy injection around $z_{\rm c}\simeq 3100$. As per our discussion in section \ref{sec:intro} (see also ref.~\cite{2019ApJ...874....4A}), the size of the baryon-photon sound horizon at the drag epoch $r_{\rm drag}$ is strongly anti-correlated with the injected energy fraction $f_\phi$. This smaller sound horizon is the main driver behind the reduction of the Hubble constant tension in these models.

The physical dark matter density ($\Omega_{\rm c}h^2$) is strongly correlated with $f_\phi$ in order to obtain the correct early integrated Sachs-Wolfe effect \cite{Vagnozzi:2021gjh}, which fixes, in part, the height of the first CMB temperature acoustic peak. This larger dark matter abundance, combined with the extra energy injected by the rolling scalar field, leads to an increase of the CMB Silk damping scales as compared to the baryon-photon sound horizon since $r_{\rm damp}/r_{\rm s}\propto H^{1/2}$ \cite{Hou:2011ec}. This is compensated in the CMB fit by increasing the values of both the scalar spectral index $n_{\rm s}$ and fluctuation amplitude $A_{\rm s}$ as compared to their \LCDM values. As a consequence of this change to the primordial spectrum of scalar perturbations, the late-time amplitude of matter fluctuations as probed by the $\sigma_8$ parameter are increased. Thus, obtaining a larger Hubble parameter within these scalar field models is correlated with an increase in the $\sigma_8$ value. This is a generic feature of models that inject energy near recombination, which might lead them to be in tension with late-time measurements of the matter power spectrum (see e.g.~ref.~\cite{Hikage:2018qbn}). We shall come back to this point in the discussion. 

\begin{figure}[t!]
  \centering
  \includegraphics[width=0.484\textwidth]{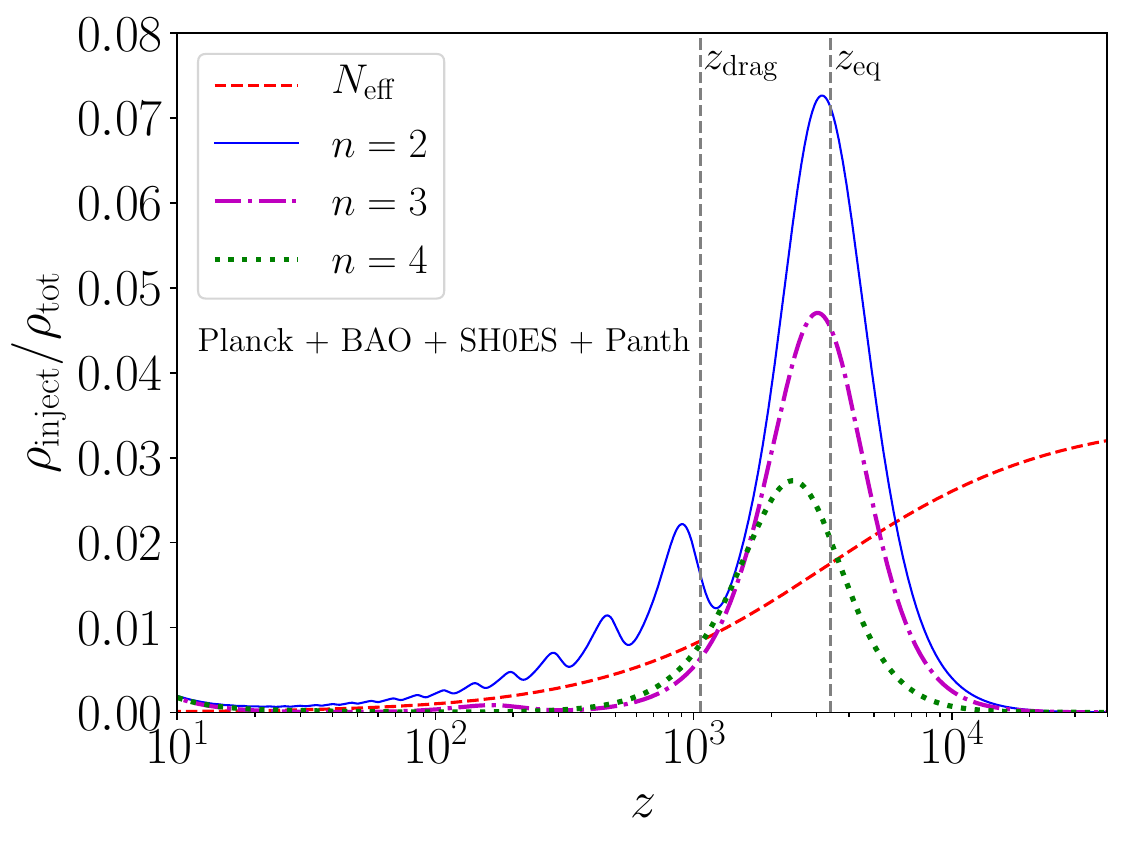}
  \includegraphics[width=0.5\textwidth]{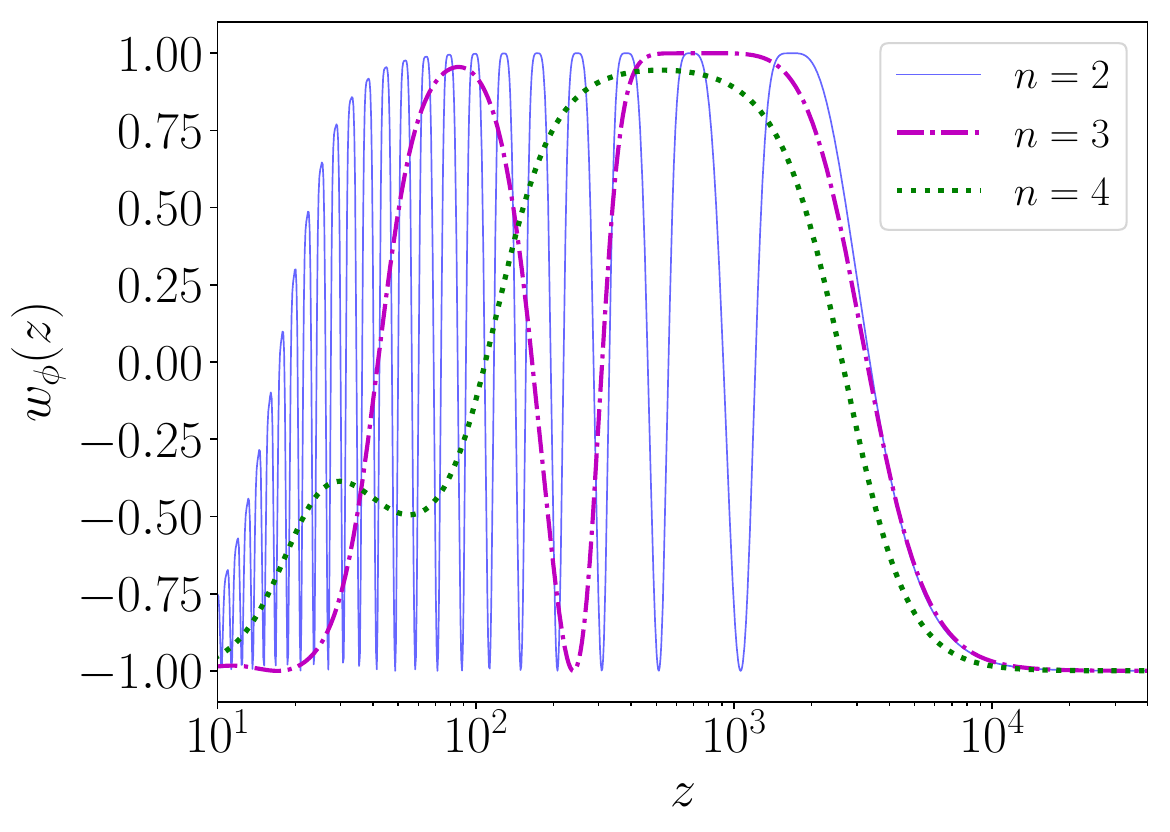}
  \caption{\emph{Left panel:} Energy injection profile for the best-fit models for each value of $n$ as a function of redshift. Results are shown here for the data combination ``Planck + BAO + SH0ES + Pantheon''. For reference, we also show the amount of energy injected as compared to standard $\Lambda$CDM for the best fit $N_{\rm eff}$ model using the same data combination (corresponding to $\Delta N_{\rm eff}=0.27$). To guide the eye, we have indicated by vertical dashed lines the matter-radiation equality and baryon drag epochs in the standard $\Lambda$CDM model. \emph{Right panel:} The scalar field equation of state as a function of redshift for each value of $n$.}
  \label{fig:energy_injection}
\end{figure}

We show in figure~\ref{fig:energy_injection} the best-fit energy-injection profile and the EoS of the scalar field for each value of $n$ we consider. We observe that, for a fixed width of the primary energy injection peak, models with lower values of $n$ allow for more energy to be injected, resulting in a smaller value of the baryon-photon sound horizon and larger Hubble constant. However, as explained in section~\ref{sec:scaling_solns}, these low-$n$ models also have strong oscillatory solutions with secondary energy injection peaks post recombination, which adversely affect the fit to CMB data. 
On the other hand, as was shown in figure~\ref{fig:emden-fowler}, models with large values of $n$ have a broader primary energy injection peak and slowly decaying tail, which means that it is difficult for these models to inject sufficient energy prior to recombination while not having significant residual energy towards low redshifts. The fit to cosmological data is thus driven by the competing effects of injecting the largest possible amount of energy in a relatively narrow window prior to recombination, while at the same time being able to get rid of this energy as quickly as possible in the post-recombination era.  In contrast with the results of section~\ref{subsec:E-F}, which were derived in a toy cosmology with $w_b = 0$, the best-fit solutions for $n=3$ and $n=4$ never achieve their asymptotic equations of state due to the effect of the cosmological constant at late times.  The salient feature, however, is the relatively large instantaneous value of $w_{\phi} \approx 1$ obtained shortly after the peak of the energy injection for all values of $n$.  This causes most of the energy density to quickly redshift away, so that the fit is relatively insensitive to the subsequent dynamics.  The speed of the transition from $w_{\phi} \simeq -1$ to $w_{\phi} \simeq 1$ determines the width of the energy injection; as shown in figure~\ref{fig:energy_injection}, this width is minimized at small $n$.

The finding that lower $n$ values provide a better fit to cosmological
data is supported by examining the individual best-fit $\chi^2$
values. We present these in table \ref{tab:best_fit_chi2} for $n=2$,
$3$, and $4$, as well as for $\Lambda$CDM\footnote{These $\chi^2$ values were obtained by running a steepest-gradient minimizer on the relevant likelihoods.}. We find that the $n=2$
model provides that best overall improvement of the fit as compared to
$\Lambda$CDM. We observe, however, that the improved fit to the SH0ES
measurement of $H_0$ is partially offset by a degradation of the
goodness-of-fit to the high-$\ell$ CMB tail and to the CMB lensing
spectrum. We note that the results shown in table
\ref{tab:best_fit_chi2} are significantly different than those
presented in ref.~\cite{Poulin:2018cxd} due to their use of a
coarse-grained effective fluid approach. As discussed in section
\ref{sec:scaling_solns} above, the coarse-grained approach is a
reasonable approximation for $n=2$ deep in the matter-dominated era
(see right panel of figure \ref{fig:energy_injection} for $z<500$),
but it fails to accurately capture the most important part of scalar
field evolution near the peak of the energy injection. The
coarse-grained approach is inaccurate for $n\geq3$ since the models
undergoes at most two oscillations before the onset of dark energy
domination, resulting in very different fits once the exact evolution
of the scalar field is taken into account. 

\begin{table}
\centering
\begin{tabular}{|c | c | c | c | c | c |}
\hline
\hline
Datasets & $\Lambda$CDM & $n=2$ & $n=3$ & $n=4$ & $N_{\text{eff}}$ \\
\hline
\hline
{\it Planck} high-$\ell$  & 2439.0  & 2440.4  & 2440.0  & 2440.5  & 2442.0 \\
{\it Planck} low-$\ell$   & 10495.8 & 10494.3 & 10494.4 & 10494.6 & 10494.5 \\
{\it Planck} lensing      & 9.5     & 10.0    & 10.2    & 10.0    & 9.6 \\
BAO - low $z$             & 1.6     & 1.6     & 1.5     & 1.5     & 2.1 \\
BAO - high $z$            & 1.8     & 1.9     & 1.9     & 1.9     & 1.9 \\
Pantheon                  & 1027.0  & 1027.0  & 1027.0  & 1027.0  & 1026.9 \\
SH0ES                     & 11.4    & 3.2     & 4.5     & 4.9     & 5.1 \\
\hline
Total $\chi^2_{\rm{min}}$ & 13986.1 & 13978.3 & 13979.5 &13980.3  & 13982.1 \\
$\Delta \chi^2_{\rm{min}}$& 0       & -7.8    & -6.6    & -5.8    & -4.0 \\
\hline
\end{tabular}
\caption{Best-fit $\chi^2$ values for each individual dataset used in our cosmological analysis. \label{tab:best_fit_chi2}}
\end{table}

\begin{table}
\centering
\begin{tabular}{|c | c | c | c |}
\hline
\hline
Parameter & $\Lambda$CDM & $n=2$ & $N_{\text{eff}}$ \\
\hline
\hline
$100\ \Omega_{\rm b} h^2$ & 2.238 (2.234) \small${}^{+0.014}_{-0.015}$ & 2.261 (2.270) \small${}^{+0.021}_{-0.020}$ & 2.254 (2.253) $\pm 0.018$ \\
\hline
$\Omega_{\rm c} h^2$ & 0.1180 (0.1182) $\pm 0.0012$ & 0.1264 (0.1284) \small${}^{+0.0044}_{-0.0043}$ & 0.1220 (0.1227) \small${}^{+0.0027}_{-0.0028}$ \\
\hline
100 $\theta_{\rm s}$ & 1.0420 (1.0419) $\pm 0.0003$ & 1.0415 (1.0414) $\pm 0.0004$ & 1.0414 (1.0413) \small$^{+0.0004}_{-0.0005}$ \\
\hline
$\tau_{\rm reio}$ & 0.074 (0.072) \small${}^{+0.013}_{-0.012}$ & 0.072 (0.074) \small${}^{+0.013}_{-0.012}$ & 0.075 (0.073) \small${}^{+0.013}_{-0.012}$ \\
\hline
$\ln (10^{10} A_{\rm s})$ & 3.079 (3.074) \small${}^{+0.024}_{-0.021}$ & 3.091 (3.097) \small${}^{+0.026}_{-0.023}$ & 3.089 (3.087) \small${}^{+0.025}_{-0.022}$ \\
\hline
$n_{\rm s}$ & $0.968\ (0.966) \pm 0.004$ & 0.978 (0.966) $\pm 0.007$ & $0.977\ (0.980)$ \small${}^{+0.006}_{-0.007}$ \\
\hline
$f_\phi$ / $\Delta N_{\text{eff}}$ & - & 0.064 (0.080) \small${}^{+0.031}_{-0.028}$ & $0.26\ (0.29) \pm 0.16$\\
\hline
$z_{\rm c}$ & - & 3040 (2902) \small${}^{+330}_{-630}$ & - \\
\hline
$\sigma_8$ & 0.819 (0.817) \small${}^{+0.009}_{-0.008}$ & 0.835 (0.840) $\pm 0.012$ & 0.831 (0.831) $\pm 0.011$ \\
\hline
$\Omega_{\rm m}$ & 0.304 (0.305) $\pm 0.007$ & 0.304 (0.304) $\pm 0.007$ & 0.299 (0.299) \small${}^{+0.007}_{-0.008}$ \\
\hline
$r_{\text{drag}}$ [Mpc] & 147.6 (147.6) $\pm 0.3$ & 143.2 (142.1) \small${}^{+2.0}_{-2.3}$ & 145.1 (144.8) $\pm 1.5$ \\
\hline
$H_0$ [km/s/Mpc] & $68.2\ (68.0) \pm 0.5$ & 70.1 (70.6) \small${}^{+1.0}_{-1.2}$ & $69.7\ (69.9) \pm 1.1$ \\
\hline
\end{tabular}
\caption{Mean values and $68\%$ confidence intervals for key cosmological parameters using the data combination  ``Planck + BAO + SH0ES + Pantheon''. The numbers in parentheses are the best-fit values for each model. \label{tab:confidence_interval}}
\end{table}

We show in table \ref{tab:confidence_interval} the mean, 1$\sigma$
uncertainty, and best fit values for the most relevant cosmological
parameters and quantities.  The $n=2$ best-fit model has the largest
Hubble constant value of all models studied in this paper, which is a
direct consequence of it having the largest energy injection fraction
(see figure \ref{fig:energy_injection}), and therefore the smallest
photon-baryon drag horizon. The physical dark matter density is also
significantly larger for the $n=2$ best-fit model, as is the amplitude
of the matter power spectrum as captured through the $\sigma_8$
parameter, for reasons explained above. We will perform a thorough
comparison between the $n=2$ model and $N_{\rm eff}$ in section
\ref{sec:N_eff} below.  

In figure \ref{fig:CMB_residuals}, we show the CMB residuals between our best-fit $\phi^{2n}$, \LCDM, and $N_{\rm eff}$ models using the data combination ``Planck + BAO + SH0ES + Pantheon'', and a reference \LCDM model fitted to ``Planck + BAO + Pantheon''  data. This allows us to isolate the impact of the SH0ES data on the residuals.  All models shown display, on average, less power in $C_\ell^{TT}$ at $\ell > 500$ compared to our reference \LCDM model. This is caused by the well-known \cite{Hou:2011ec} competition between having a large enough Hubble constant to fit the SH0ES measurement while ensuring that there is enough power in the CMB temperature Silk damping tail. We note, however, that the variations between the different models shown are typically smaller than the size of the error bars.

\begin{figure}[t!]
  \centering
  \includegraphics[width=0.495\textwidth]{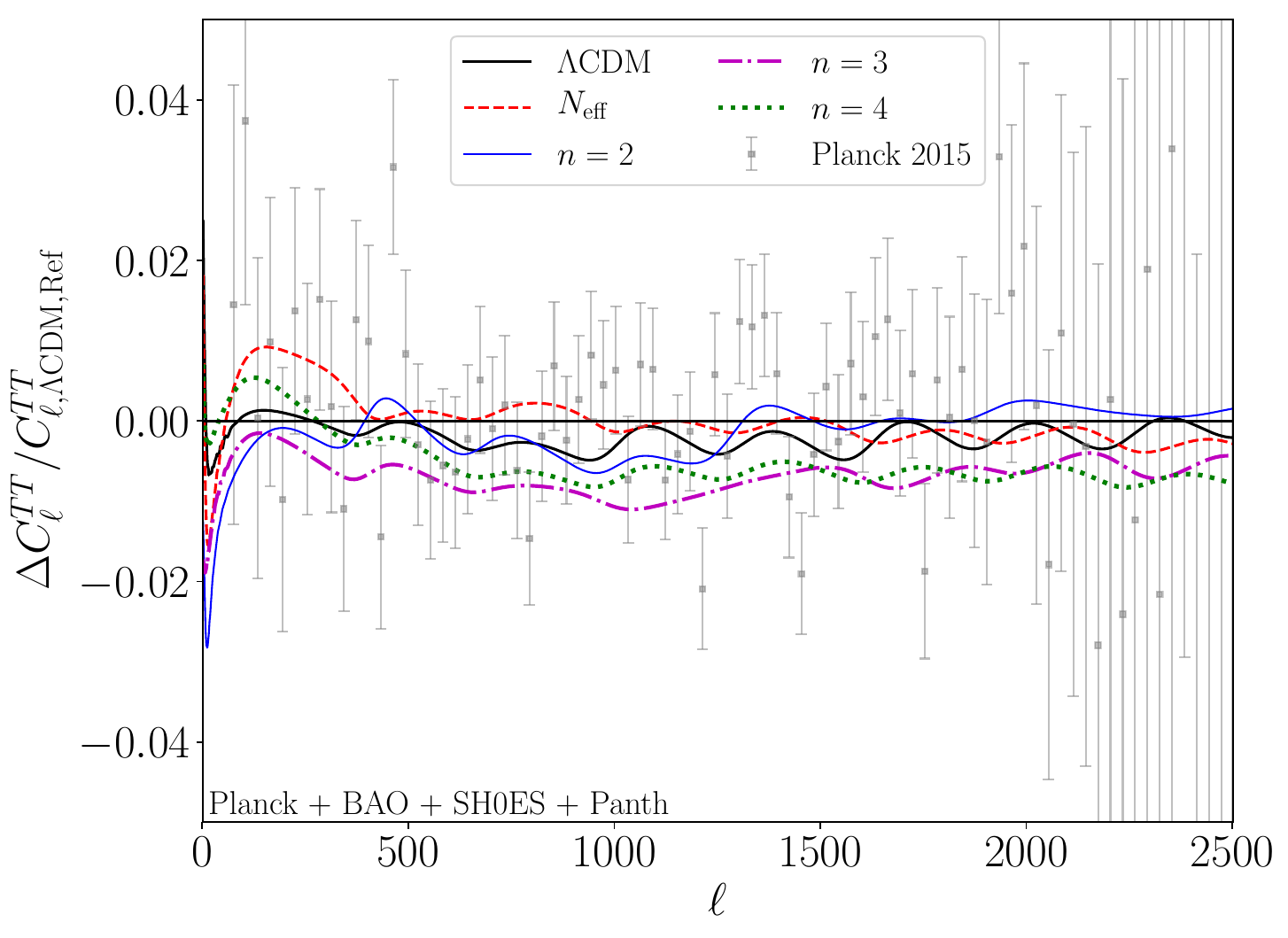}
  \includegraphics[width=0.495\textwidth]{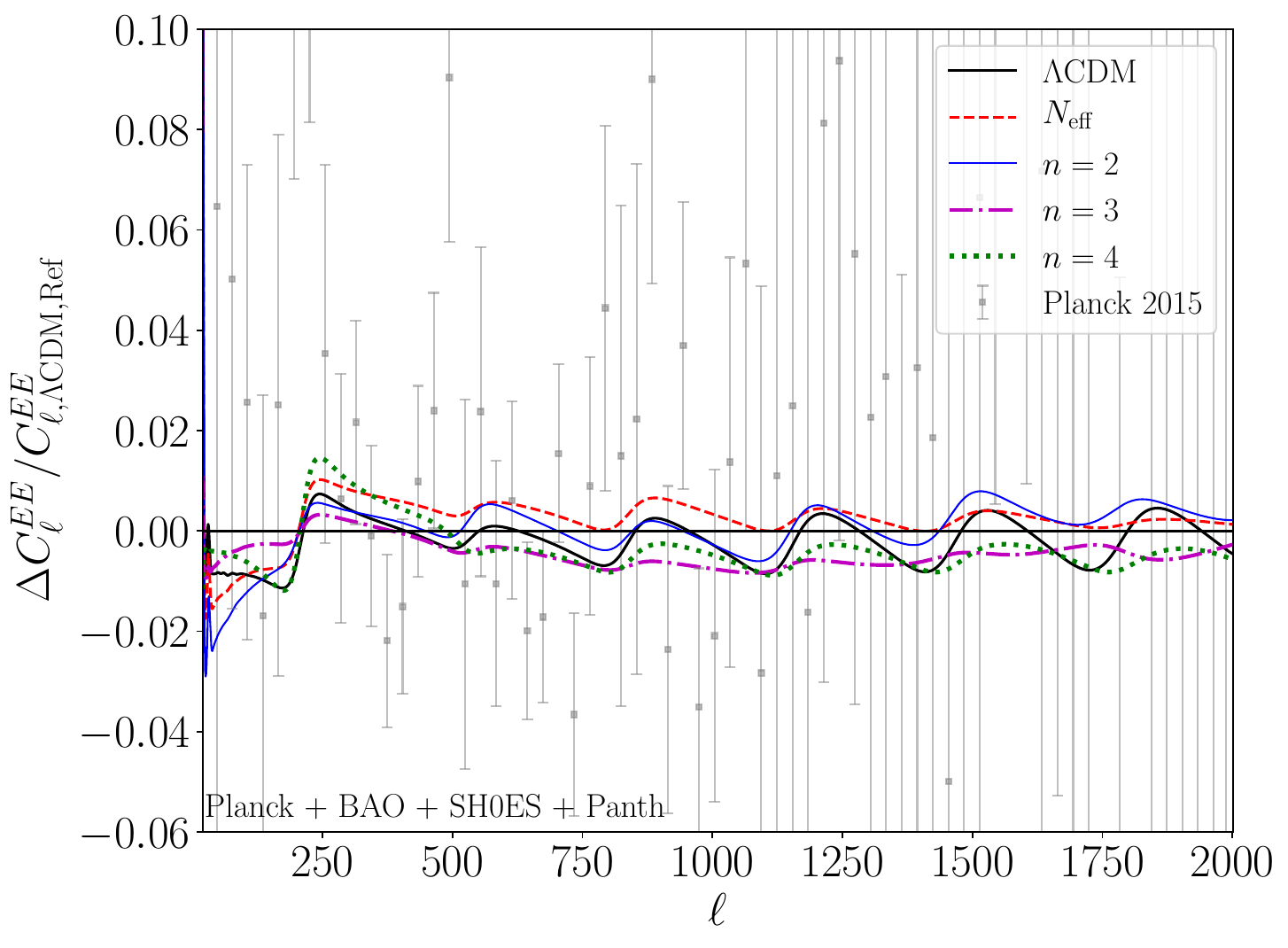}
  \caption{High-$\ell$ CMB residuals between the ``Planck + BAO +
  SH0ES + Pantheon'' best fit models for \LCDM, $N_{\rm eff}$, and $\phi^{2n}$, and a best-fit \LCDM reference model obtained using the data combination ``Planck + BAO + Pantheon''. The left panel shows the temperature residuals, while the right panel shows the E-mode polarization residuals. }
  \label{fig:CMB_residuals}
\end{figure}

\begin{figure}[b!]
  \centering
  \includegraphics[width=0.495\textwidth]{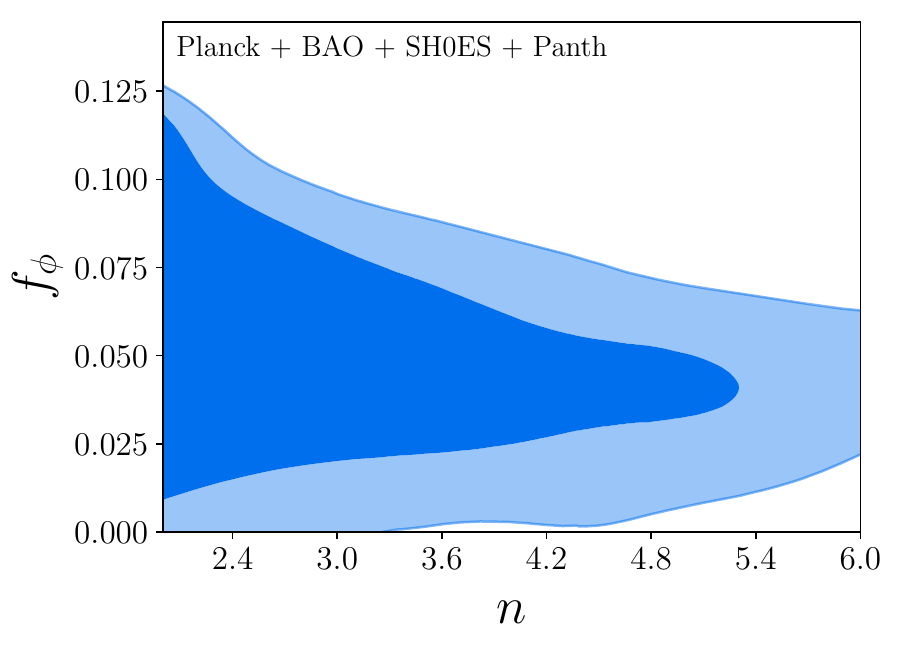}
  \includegraphics[width=0.49\textwidth]{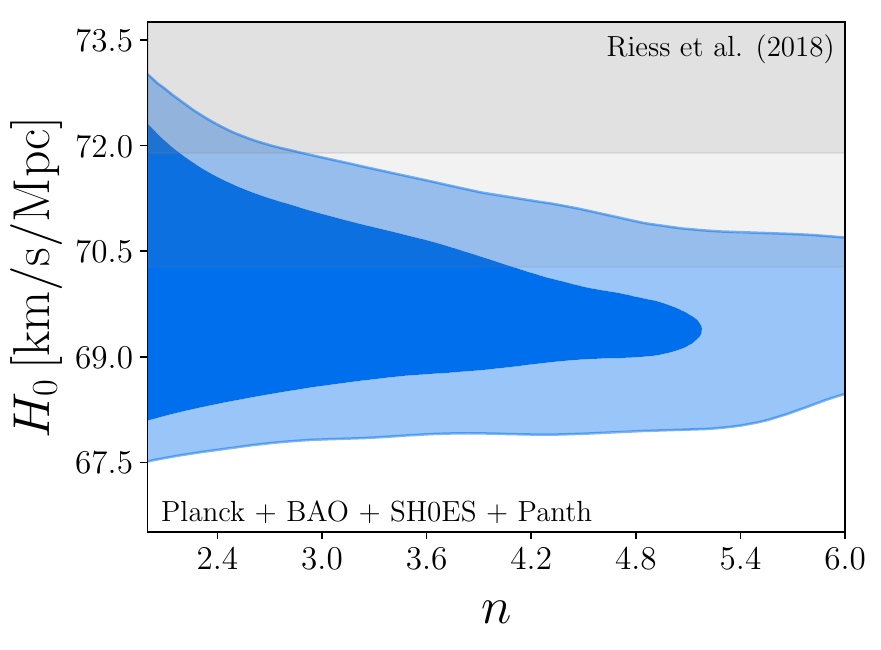}
  \caption{ Marginalized posterior in the $f_\phi$-$n$ plane (left panel), and the $H_0$-$n$ plane (right panel). The gray band shows the SH0ES measurement \cite{Riess:2018byc}. Results are shown here for the data combination ``Planck + BAO + SH0ES + Pantheon.'' Clearly, larger values of the Hubble constant require lower values of $n$.}
  \label{fig:free_n}
\end{figure}

\subsection{Results: Promoting \texorpdfstring{$n$}{n} to a free parameter }
So far, we have considered only fixed integer values of the power-law
index $n$ parameterizing the scalar potential. This was effectively
restricting the behavior of the energy injection to have a specific
shape specified by our choice of $n$. In this section, we relax the
assumption of fixed integer values of $n$ and allow the power-law
index to float freely in the fit to cosmological data (writing our
  potential as $V\propto |\phi|^{2n}$), hence exploring a broader range
  of energy injection shapes and scalar field behaviors. While we
  adopt this simple phenomenological point-of-view, we note that
  models with noninteger $n$ values may be problematic from a
  model-building perspective as they typically correspond to nonlocal theories (see e.g.~ref.~\cite{Ignatyuk:2023psc}). We consider a flat prior on $n\in[2,6]$.
  The main results of this analysis are shown in figure
  \ref{fig:free_n}.  As discussed above, the data tends to favor lower
  values of the index $n$ since they allow a larger peak energy
  injection fraction. The right panel clearly illustrates that larger
  Hubble constant values are more likely for the lowest possible index
  $n$, which, given our choice of priors, corresponds to $n\sim2$
  (that is, $V\propto\phi^4$).  

Figure \ref{fig:free_n} suggests that one should consider even lower values of $n$ in order to fit the large Hubble constant from the SH0ES measurement, despite these solutions being increasingly oscillatory. To explore this possibility, we extend our prior range on $n$ to include values all the way down to $n=1.5$, a compromise between exploring low values of $n$ while at the same time ensuring that the oscillatory solutions are tractable numerically. We find that it is indeed possible to obtain larger values of the Hubble constant (our best for with $n< 2$ has $H_0=71.74$ km/s/Mpc for $n=1.89$), but at the price of significantly degrading the fit to the CMB damping tail. Overall, we find that none of the models with $n< 2$ provides a better global fit to cosmological data than the $n=2$ model presented in tables \ref{tab:best_fit_chi2} and \ref{tab:confidence_interval}.

\subsection{Results: Comparison with \texorpdfstring{$N_{\rm eff}$}{Neff}}\label{sec:N_eff}

It is informative to compare the scalar field models with the more standard $N_{\rm eff}$ extension to the $\Lambda$CDM model. The amount of energy injected by the best-fit $N_{\rm eff}$ model is shown in figure \ref{fig:energy_injection} (corresponding to $\Delta N_{\rm eff}=0.27$). The addition of relativistic species leads to a broadband energy injection, in contrast to the relatively narrow energy injection peaks of the $n=2$ model. Despite these differences, we see in table \ref{tab:best_fit_chi2} that the fit to the high-$\ell$ CMB data is similar to the $n=2$ scalar field model. We note however that these two fits are significantly worse than that achieved in the reference \LCDM model fitted only to  ``Planck + BAO  + Pantheon.''  The main differences between the two models is the physical amount of dark matter required to obtain the correct magnitude of the early ISW effect. Since the $N_{\rm eff}$ model injects less relativistic energy near matter-radiation equality than the $n=2$ model, a smaller amount of dark matter is required to leave the early ISW effect invariant.  In turn, this leads to a smaller $\Omega_{\rm m}$ than in $n=2$, which worsens the fit to late-time probes such as BAO. Thus, while $N_{\rm eff}$ models can accommodate larger values of the Hubble constant without degrading the global fit, they do so by compromising the fit to the low-$z$ BAO data as shown in the last column of table \ref{tab:best_fit_chi2}. 

\begin{figure}[t!]
  \centering
  \includegraphics[width=\textwidth]{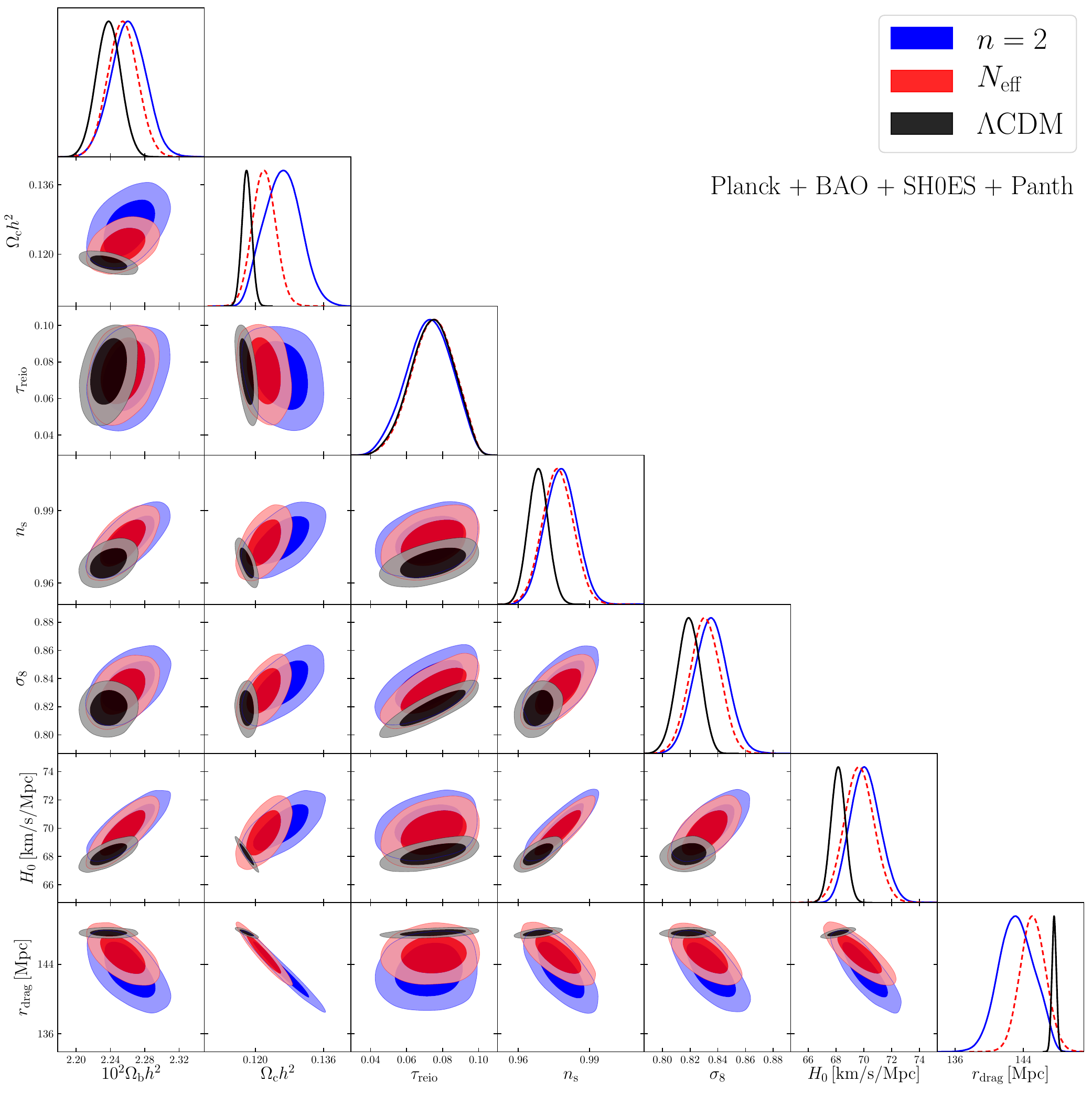}
  \caption{Comparison between the $n=2$ model, the $N_{\rm eff}$ extension of the standard cosmological model, and plain $\Lambda$CDM. All posterior distributions shown here were obtained using the data combination ``Planck + BAO + SH0ES + Pantheon''. }
  \label{fig:triangle_comp_Neff_phi4}
\end{figure}

By comparison, the $n=2$ scalar field model has a similar fit to the CMB high-$\ell$ tail as the $N_{\rm eff}$ model, while at the same time typically allowing for larger Hubble constant and better fit to BAO data. We must however keep in mind that the $\phi^{2n}$ models introduce two free parameters ($f_\phi$, $z_{\rm c}$) to the fit, compared to the single parameter in $N_{\rm eff}$ models. We can use the Aikake information criterion (AIC)  \cite{AIC}, which can be defined as $\Delta {\rm AIC} = \Delta\chi^2_{\rm min} + 2\Delta k$, where $\Delta k$ is the difference in the number of free parameters between the two models considered. Using the $\Lambda$CDM fit to the ``Planck + BAO + SH0ES + Pantheon'' dataset as our reference model, we obtain $\Delta {\rm AIC}_{n=2} = -2.9$, while $\Delta{\rm AIC}_{N_{\rm eff}} = -2.4$, indicating that the $n=2$ scalar field model provides a slightly better fit to the data as compared to $N_{\rm eff}$, even after accounting for the extra free parameter. In comparison, the $n=3$ and the $n=4$ models provide a worse global fit to the data than $N_{\rm eff}$ after accounting for the different numbers of free parameters. This reinforces the fact that models with lower values of the power-law index $n$ are favored by the data.   On the other hand, $N_{\rm eff}$ models are theoretically somewhat simpler.  UV completions of $\phi^{2n}$ models must explain not only the absence of additional self-couplings, but also the surprising coincidence between the timing of the energy injection and the redshift of matter-radiation equality demanded by the fit to CMB data. 

A quantitative comparison between the $n=2$ scalar field model and the $N_{\rm eff}$ extension of $\Lambda$CDM is shown in figure \ref{fig:triangle_comp_Neff_phi4}. We first observe that the scalar field model generally allows for smaller values of the baryon drag horizon $r_{\rm drag}$ than the $N_{\rm eff}$ model. Naively, following the argument presented in ref.~\cite{2019ApJ...874....4A}, this should allow the scalar field model to accommodate a larger Hubble constant as compared to $N_{\rm eff}$. Instead, we see that the two models have fairly similar posterior distribution for $H_0$, with the scalar field model allowing only slightly larger values of the expansion rate. This difference between our naive expectation and what is shown in figure \ref{fig:triangle_comp_Neff_phi4} is caused by the different amount of dark matter needed in each model, as described above. This larger dark matter abundance pulls down on the Hubble constant value in the cosmological fit, partially compensating for the increase coming from a smaller baryon sound horizon. The scalar field models with $\phi^{2n}$ potentials are thus less efficient at turning a given reduction of the sound horizon into an increase of $H_0$, as compared to $N_{\rm eff}$. 

Both $n=2$ and $N_{\rm eff}$ models have larger values of $\sigma_8$ which are largely caused by the change to the primordial spectrum of fluctuations ($A_{\rm s}$ and $n_{\rm s}$) necessary to bring the CMB damping tail in agreement with the data. As discussed in section \ref{sec:res_fixed_n}, this is a generic feature of models that inject energy near the epoch of recombination. The effect is larger for the $n=2$ model since more energy is injected overall, leading to a slightly larger value of $\sigma_8$ as compared to $N_{\rm eff}$. 

\section{Discussion}\label{sec:disc}
\begin{figure}[t!]
  \centering
  \includegraphics[width=0.6\textwidth]{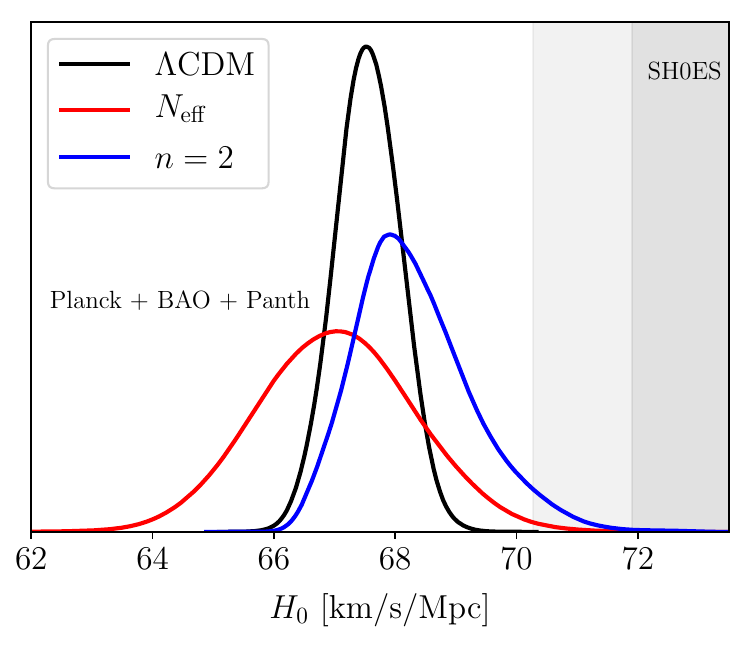}
  \caption{Normalized $H_0$ posteriors obtained using the data combination ``Planck + BAO + Pantheon'', that is, without including the local Hubble constant measurement from ref.~\cite{Riess:2018byc}. }
  \label{fig:H0_tension}
\end{figure}

We have seen that a scalar field with $V\propto \phi^{2n}$ leads to an energy injection which is localized in time, potentially alleviating the tension between local measurements of the Hubble constant and that inferred from CMB observations. One important worry is the slight degradation of the fit to CMB data when Planck and SH0ES measurements are analyzed simultaneously.  Does this indicate that Planck data alone disfavors the energy injection brought by the rolling scalar field? To answer this question, we redo our MCMC analyses for the $\phi^{2n}$ models but this time without the SH0ES likelihood. In figure \ref{fig:H0_tension} we show the normalized posterior of $H_0$ resulting from these analyses for \LCDM, $N_{\rm eff}$, and $n=2$. While both the $n=2$ and $N_{\rm eff}$ posteriors are
wider than \LCDM, that of $n=2$ is peaked at marginally higher values of
$H_0$ than either alternative model, with $N_{\rm eff}$ actually peaking below \LCDM. This suggests that the energy injection brought by the evolving scalar field can naturally accommodate larger Hubble constants in the CMB fit, even without prior information from late-time measurements. While the $n=2$ model does not entirely remove the $H_0$ tension, it can significantly reduce it to the $\sim2\sigma$ level.  Given the longer tail of the posterior distribution, future local measurements of $H_0$ which reduce the error bars could more strongly distinguish the scalar field model from $N_{\rm eff}$.

\begin{figure}[t!]
  \centering
  \includegraphics[width=0.495\textwidth]{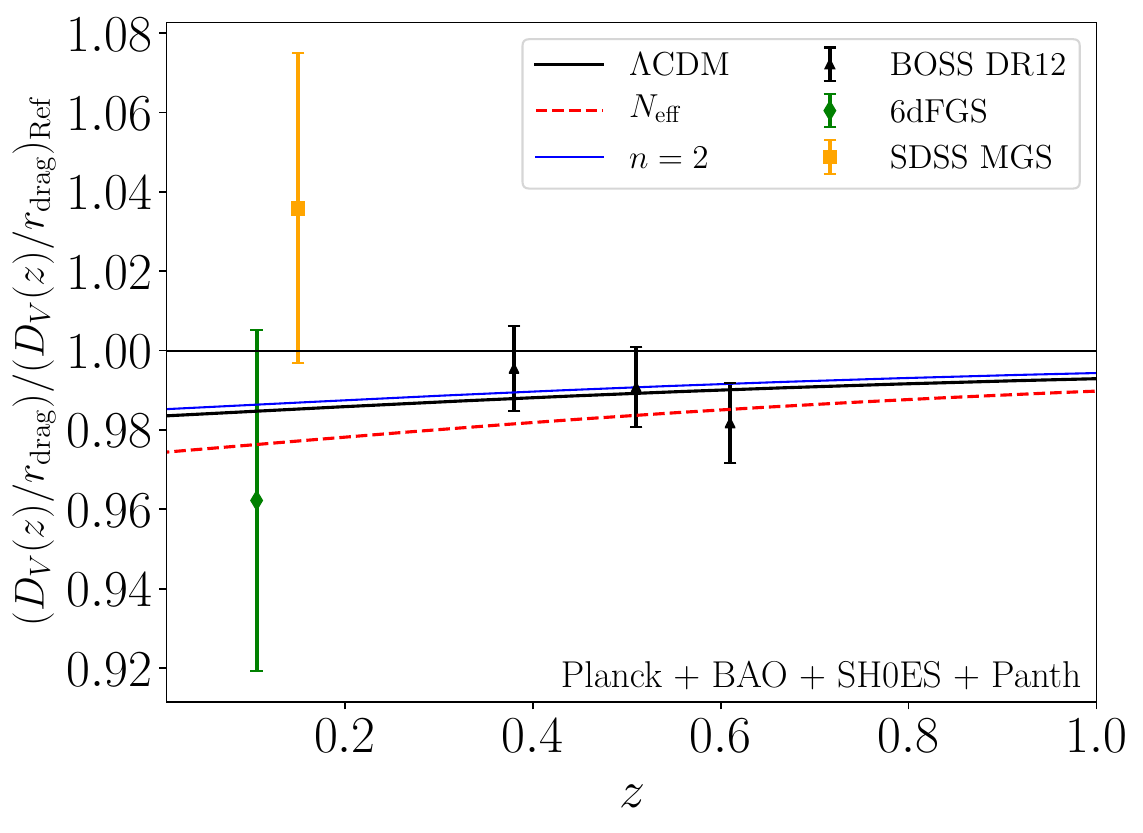}
  \includegraphics[width=0.495\textwidth]{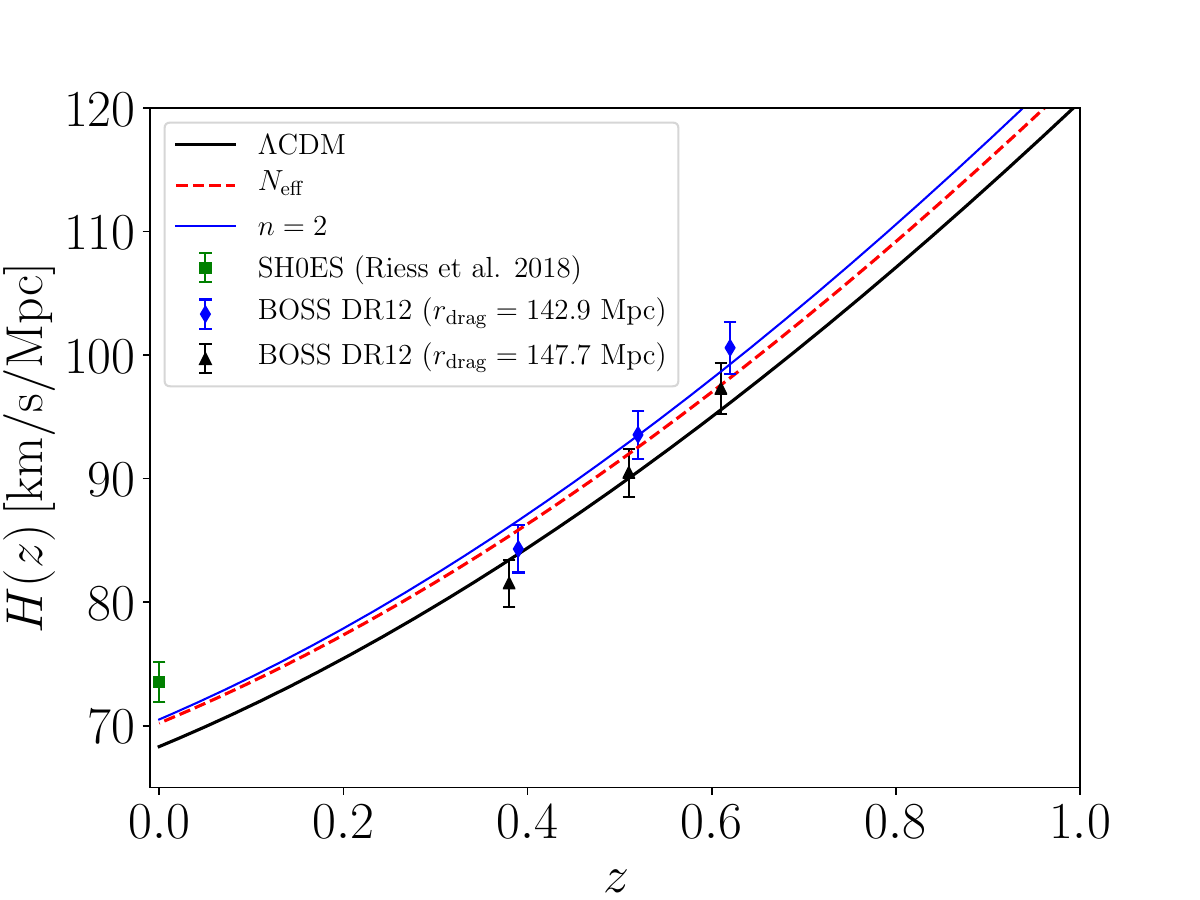}
  \caption{\emph{Left panel:} The BAO distance ladder as expressed through the ratio $D_V(z)/r_{\rm drag}$ normalized to the best-fit ``Planck + BAO + Pantheon'' \LCDM reference model. \emph{Right panel:} Hubble expansion history for $z<1$. We show the $H_0$ measurement from ref.~\cite{Riess:2018byc}, as well as two sets of line-of-sight BOSS DR12 BAO measurements \cite{Alam:2016hwk}: one calibrated to the sound horizon of the best-fit \LCDM model (black triangles), and the other calibrated to that of the best-fit $n=2$ model (blue diamonds).  All curves shown in this figure are best-fit models to the data combination ``Planck + BAO + SH0ES + Pantheon''.}
  \label{fig:BAO_Hubble}
\end{figure}

It is also interesting to examine the impact of the $\phi^{2n}$ models on late-time cosmology. We show in figure \ref{fig:BAO_Hubble} the changes to the cosmology at $z<1$ for the different models studied in this paper. The left panel shows the fit to the BAO measurements used in our analyses. For clarity, we illustrate the quantity $D_V(z)/r_{\rm drag}$ for the best-fit parameters of the different models considered, normalized by that of the ``Planck + BAO + Pantheon''  \LCDM model. We see that all models have lower values of $D_V(z)/r_{\rm drag}$ over the redshift range shown.  Expanding around $z = 0$,
\begin{align}
  \frac{D_V (z)}{D_{V,{\rm fid}} (z)}
  &\approx
  \frac{H_{0,{\rm fid}}}{H_0} \left[1 + \left(\Omega_{\rm m, fid} - \Omega_{\rm m}\right) z + \mathcal{O}(z^2)\right].
\end{align}
In all of the models considered, $\Omega_{\rm m}$ is decreased relative to the best-fit ``Planck + BAO + Pantheon'' $\Lambda$CDM reference model, despite the relative increase in $\Omega_{\rm c} h^2$.  This produces a positive slope for the $D_V$ ratio at late times, guaranteeing that those models which fit the high-$z$ BAO data at redshift 0.5 will be in mild tension with the MGS measurement at lower redshift. Since $N_{\rm eff}$ has the smallest $\Omega_{\rm m}$ of all models shown, this effect is more severe for this model.

The right panel of figure \ref{fig:BAO_Hubble}  shows the Hubble expansion history at $z<1$ for our best-fit $n=2$ model, and compares it to that of the best-fit \LCDM and $N_{\rm eff}$ models (for the data combination ``Planck + BAO + SH0ES + Pantheon''). To illustrate how the scalar field model helps reconciling the CMB, BAO, and SH0ES measurements of the expansion rate, we also show two sets of line-of-sight BOSS DR12 BAO measurements \cite{Alam:2016hwk}. The first set is calibrated using the sound horizon from $n=2$ scalar field fit to the CMB ($r_{\rm drag} =142.9$ Mpc), while the second uses the sound horizon from the best-fit \LCDM model ($r_{\rm drag} = 147.7$ Mpc). We observe in the right panel that the $n=2$ model fits the properly calibrated BAO data points nearly as well as \LCDM, while at the same time providing a much better fit to the SH0ES measurement.

\begin{figure}[t!]
  \centering
  \includegraphics[width=0.6\textwidth]{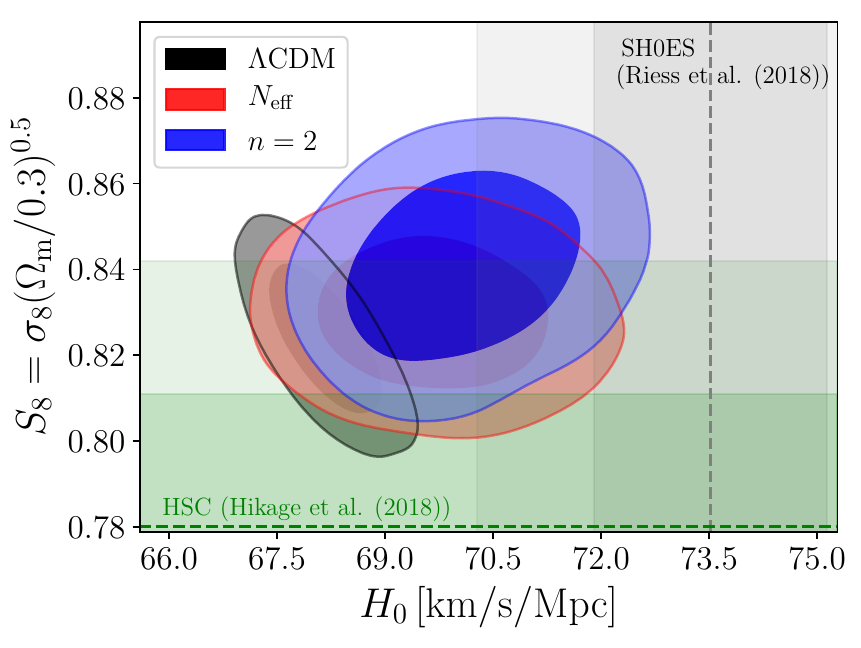}
  \caption{Marginalized posterior distributions in the $S_8$-$H_0$ plane obtained using the data combination ``Planck + BAO + SH0ES + Pantheon''. The gray vertical band shows the local Hubble constant measurement from ref.~\cite{Riess:2018byc}, while the horizontal green band shows the $S_8$ measurement from ref.~\cite{Hikage:2018qbn}. }
  \label{fig:S8_H0}
\end{figure}

As discussed in section \ref{sec:res_fixed_n}, rolling scalar field
models that inject energy in the period just prior to recombination
require larger values of the scalar spectral index and amplitude of primordial fluctuations. This causes the late-time amplitude of matter fluctuations to be larger than their \LCDM counterparts. This is potentially problematic since several measurements of the quantity $S_8\equiv \sigma_8\sqrt{\Omega_{\rm m}/0.3}$ have returned slightly lower values than the \LCDM expectation based on its CMB fit.  We illustrate in figure \ref{fig:S8_H0} the joint marginalized posterior of $S_8$ and $H_0$ for both the $n=2$ and $N_{\rm eff}$ models. Since the $n=2$ scalar field model has slightly larger values of both $\sigma_8$ and $\Omega_{\rm m}$ as compared to $N_{\rm eff}$, we find that it worsens the tension with the value of $S_8$ measured from weak lensing observations (e.g.~ref.~\cite{Hikage:2018qbn}). We note however that the robustness of this $S_8$ tension is still debated (see e.g.~refs.~\cite{Efstathiou:2017rgv,Nunes:2021ipq}) and more work is necessary to establish whether this is a problem for the kind of models considered here.

Another possible issue with these scalar field models is that the larger values of the Hubble constant is correlated with a larger reionization optical depth $\tau_{\rm reio}$, as can be seen in figure \ref{fig:triangle_comp_Neff_phi4}. Such large values of $\tau_{\rm reio}$ could require new high-redshift sources in order to reionize the Universe at a higher redshift \cite{Robertson:2015uda}. More recent analyses (see e.g.~ref.~\cite{Smith:2022hwi}) have found lower values of $\tau_{\rm reio}$ than we find here, although they are somewhat still larger than in \LCDM.

Up to now, we have considered pure monomial potentials, $V \propto
\phi^{2n}$, in which the more relevant $\phi^{2 k}$ terms with $k < n$
have been set to zero.  In realistic UV completions, however, one
expects that these terms would be present at some order, and their
absence would then reflect some degree of tuning in the model.  For a
sufficiently small mass term, the dominant effect is that the residual
energy density in the scalar field comes to behave like an additional
dark matter component once the scalar field amplitude has decayed such
that $\langle m^2 \phi^2 \rangle \sim \rho_{\phi}$.  At larger
amplitudes, the field behaves as described in
section~\ref{sec:scaling_solns}.  This additional dark matter slightly
modifies the distance-redshift relationship, in particular decreasing
the distance to the surface of last scattering.  As long as the dark
matter energy density is increased by less than one part in
$\mathcal{O} (10^{-3})$, the effects are within our stated error bars
and the above results will hold. For our best fit $n=2$ model, this
corresponds to a mass below $\sim 10^{-4} {\rm eV}^4 / M_{\rm pl}^2$,
representing at most a 1\% tuning given the size of the quartic
coupling.  We have neglected to include such a term for simplicity,
and including it could actually improve the fit.  While the best fit
model has more dark matter than is found in \LCDM, the increase
in $H_0$ causes $\Omega_m$ to be slightly smaller, representing an
increase in the redshift of dark energy domination.  Including a small
amount of additional dark matter in the late universe relative to its
abundance near matter-radiation equality would then improve the fit to
low-z BAO and supernovae data, at the cost of an increase in the
required numerical precision to track the rapid field oscillations.
We leave a detailed study to future work.

\section{Conclusions}
\label{sec:conc}
We see that scalar fields are good candidates for localized energy
injection in the early universe that can in principle produce better
combined fits to the CMB, BAO, and SH0ES than, \emph{e.g.}, $N_{\rm eff}$. We have considered models in which the scalar field spends a long time, spends a long time, spends a long lonely, lonely, lonely, lonely, lonely time frozen before beginning to roll near matter-radiation equality.  In addition:

\begin{itemize}
\item We have classified the scalar field solutions for $\phi^{2n}$
potentials and highlighted a class of rolling solutions whose equation
of state asymptotes to a constant value.

\item Oscillating solutions exist only for $n\leq3$ in a
matter-dominated universe.
\item {A coarse-grained description of the evolution of the scalar field is insufficient to accurately describe the cosmological evolution near the maximum of the energy injection. }

\item These models predict larger values of $\sigma_8$. Improvements in this measurement as well as other, more
direct measurements of the matter power spectrum will help test
these solutions in the future. We expect these models to be
distinguishable at large $k$.

\item A weakness of the model as it stands is the requirement that
thawing occurs very close to matter-radiation equality at $T\sim$ eV.  A more complete model
would also explain this, presumably through a triggering of the onset of rolling
by the change in the background evolution (see e.g.~refs.~\cite{Umilta:2015cta,Ballardini:2016cvy,Sakstein:2019fmf}). 

\item Our analysis requires a modest tuning.  A small mass term which does not
affect the physics around the energy injection might improve the fit to low-$z$ BAO data.
It might also be interesting to search for models in which the absence of a mass term is natural.

\item In 2019, $N_{\rm eff}$ might still be a more compelling model since
there is neither fine tuning nor a coincidence problem. However, if
future measurements of the Hubble
parameter have smaller error bars with the same central value, scalar field models could become favored as they
allow larger values for $H_0$.

\end{itemize}

\noindent Clearly the Hubble tension is one of the most intriguing
discrepancies in cosmology today. Improved measurements
of the CMB, $H(z)$, BAO at various redshifts, and the matter power
spectrum (at both $\sigma_8$ and higher $k$), will ultimately give us
greater insight into the existence of physics beyond $\Lambda$CDM.

\acknowledgments
We thank Daniel Grin, Julian Mu\~noz, Mustafa Amin, Manuel Buen-Abad,
Martin Schmaltz, Daniel Eisenstein, and Vivian Poulin for useful
discussions. L.R.~wishes to thank Adam Riess for drawing our attention to
ref.~\cite{Poulin:2018cxd} and Matthias Steinmetz for an invitation to
the symposium  ``The Hubble constant controversy: Status,
implications, and solutions'', which inspired this work. D.P. wishes
to thank Boston University for hospitality during completion of this
work. We thank the Institut Henri Poincar\'e, where part of this work
was carried out, for their hospitality. F.-Y. C.-R. and D.P.~acknowledge the
support of the National Aeronautical and Space Administration (NASA)
ATP grant NNX16AI12G at Harvard University.  D.P.~is also supported by DOE
grants DE-SC-0013607 and DE-SC-0010010.  L.R.~is supported by NSF grant PHY-1620806, the Kavli
Foundation grant ``Kavli Dream Team,'' the Simons Fellows Program, the
Guggenheim Foundation, and an IHES CARMIN fellowship. The work of PA is supported by the NSF grants
PHY-0855591 and PHY-1216270. 
PA thanks the Galileo Galilei
Institute for Theoretical Physics for their hospitality and
the INFN for partial support during the completion of
this work.
The computations
in this paper were run on the Odyssey cluster supported by the FAS
Division of Science, Research Computing Group at Harvard University.

\appendix
\section{Large Backreaction}
The rolling solutions we have emphasized in section~\ref{subsec:E-F} may be somewhat surprising, given intuition derived from scalar fields evolving in non-expanding backgrounds.  In the $M_{\rm pl} \rightarrow \infty$ limit, there is no Hubble friction to damp oscillations, and so the rolling solutions do not exist.  Furthermore, familiar examples of cosmological scalar fields often display oscillatory behavior even when initially over-damped by Hubble friction, such as the inflaton at the end of inflation.  In this appendix we generalize the results of the main paper, which were derived working to zeroth order in the gravitational back-reaction, to the case where the scalar field dominates the energy
density of the universe.

In this case $\phi^{2n}$ potentials have oscillating solutions for all
values of $n>0$. We can check self-consistency of these oscillatory solutions. 
If we assume an oscillatory solution, then the total energy
density has equation of state $w_{\rm osc}$, so we can use the
Emden-Fowler
analysis above 
with the identification $w_b = w_{\rm osc}$. This implies that for  
 \begin{align}
  n &< \frac{2}{1-w_{\rm osc}}, 
  \qquad
  \rho_{\rm total} = \rho_\phi,
\end{align}
only oscillatory solutions exist,
which is always true since $w_{\rm osc} = (n-1)/(n+1)$.  

\begin{figure}[t]
  \centering
  \includegraphics[width=0.5\textwidth]{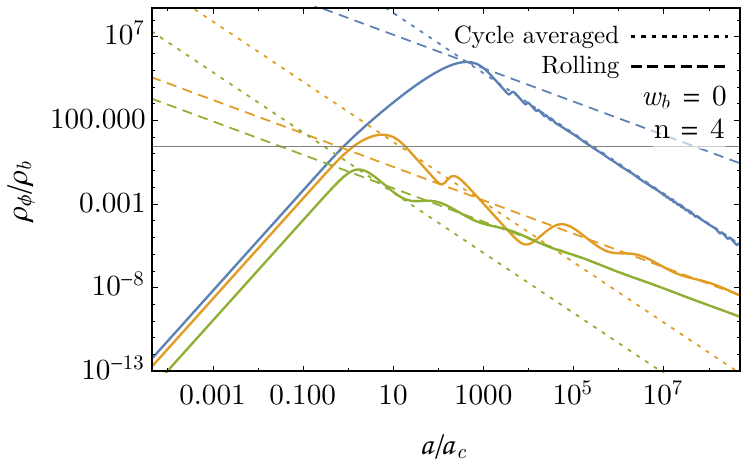}
  \includegraphics[width=0.46\textwidth]{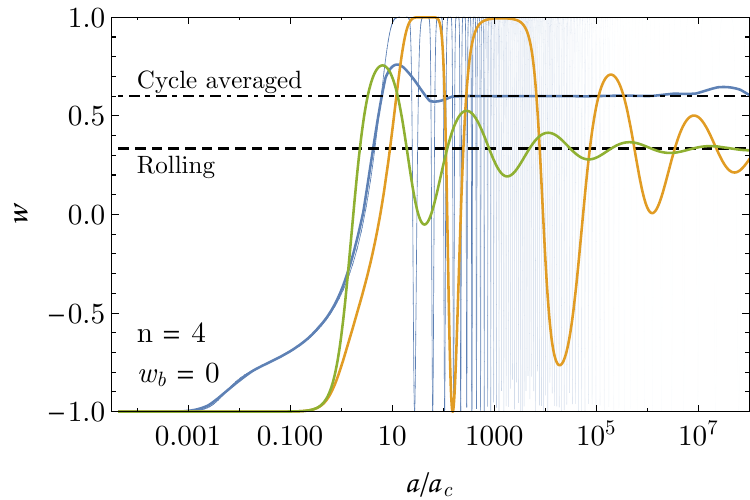}
  \caption{Evolution of the energy density (left) and the equation of
  state of the scalar field (right) as a function of the scale factor. Here, we compare the cases where the 
 scalar field eventually dominates the energy density of
the universe (blue and orange solid lines), with that in which it is always subdominant (solid green). The potential was chosen to be $V\propto \phi^{2n}$ with $n = 4$. }
  \label{fig:domination}
\end{figure}

We show an example of
such a solution for $n=4$ in figure~\ref{fig:domination}, where we take a toy matter-dominated cosmology (with $w_b = 0$).  Here the Hubble constant is determined consistently including both the $\rho_b$ and $\rho_{\phi}$ contributions.  Note that this behavior is not used for our main results,
since the fit to CMB data requires that the scalar field energy density is always subdominant. 

 We display three different choices of initial condition for the scalar field, corresponding to increasing peak energy fractions.  Along the curve in green, the scalar field energy density is always subdominant to the background energy density, and the curve quickly asymptotes to the equation of state derived in section~\ref{subsec:E-F} for rolling solution with $w_{\phi} = 1/3$.  Increasing the peak energy density beyond $\rho_b$, the solution becomes oscillatory, with a cycle-averaged equation of state given by $w_{\rm osc}$, as demonstrated by the blue and orange curves.  Eventually, the scalar field energy density redshifts enough that it becomes subdominant once again, and the field recovers the rolling solution.  This is apparent along the orange curve, which is notably kinked near $a/a_c \sim 10^5$.  It is interesting to note that this transition does not take place immediately, but rather long after the scalar field has become subdominant to the background.

\section{Beyond monomial potentials}
The asymptotic behavior of scalar fields studied in
section~\ref{sec:scaling_solns} is not unique to the monomial
potentials considered there. Indeed, any potential which asymptotes to the monomial
potential at small field values will exhibit the same asymptotic behavior.
In this section we compare our results with the early dark energy
fluid model of Ref.~\cite{Poulin:2018cxd}. This fluid model aims to
approximate~\cite{Poulin:2018dzj} a scalar field evolving on a cosine
potential, $V \propto (1-\cos(\phi/f))^n$, which we also consider
here. Of course when $\phi \ll f$, this is well approximated by $V\sim
\phi^{2n}$. 

We want to disentangle two possible explanations for differences
between our results and those of Ref.~\cite{Poulin:2018cxd}. 
First, the shape of the potential $\phi^{2n}$ is
different from the $(1-\cos(\phi/f))^n$ at large field values
Second, the smooth fluid approximation considered in Ref.~\cite{Poulin:2018cxd}
fails to capture the dynamics of an oscillating scalar field near
the peak of the energy injection.
We present a few examples to make these distinctions concrete.

\begin{figure}[tp]
  \centering
  \includegraphics[width=0.43\textwidth]{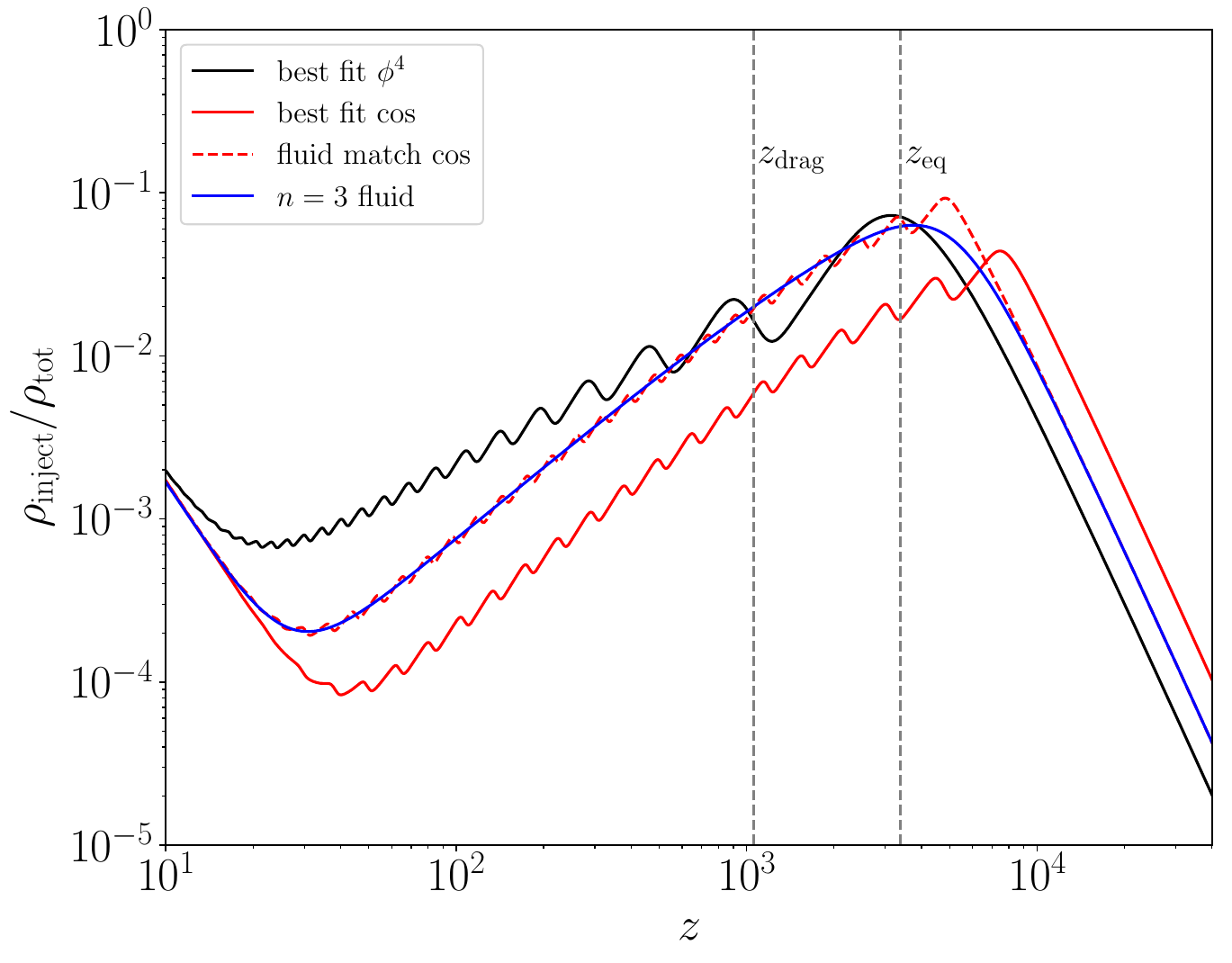}
  \qquad
  \includegraphics[width=0.45\textwidth]{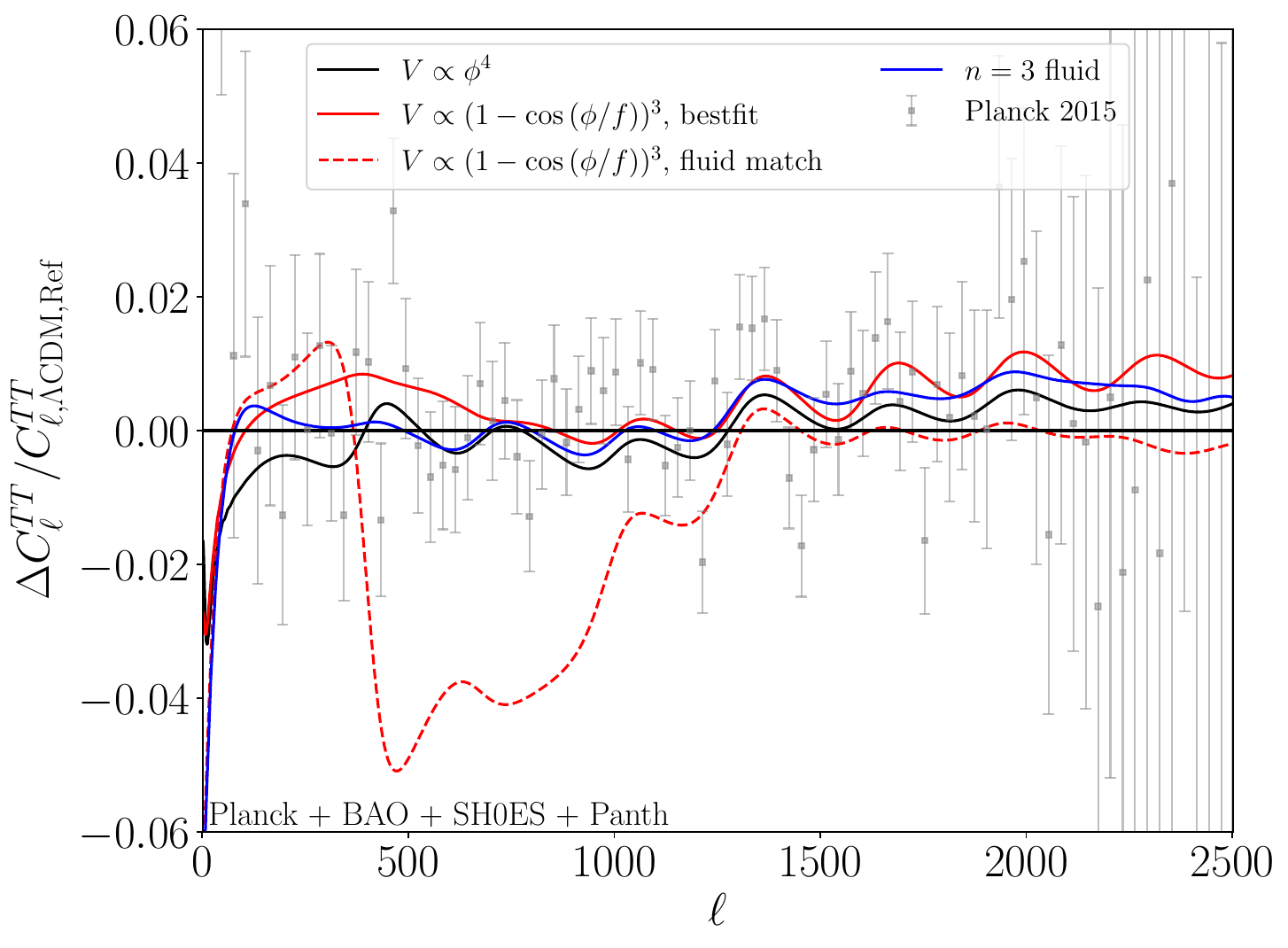}
  \caption{Energy injection profile (left) and CMB temperature residuals (right) for 4 different models. Here, we compare two different cosine potentials and a fluid model with our best fit $\phi^4$ model.}
  \label{fig:comparecos}
\end{figure}

To select benchmarks, we pick the best fit case within our models,
$n=2$ ($V\propto \phi^4$) and that of the fluid models in
Ref.~\cite{Poulin:2018cxd},
for which $n=3$. We label this the ``n=3 fluid'' model in
figure~\ref{fig:comparecos}.  For comparison, we show results for a cosine model. 
As the fluid model was designed to reproduce the results
of the cosine, we fix $n=3$ for the cosine as well.

The remaining parameters in the cosine model are the coefficient of the
potential, the initial field value, and the field range $f$. Two of
these parameters control the height and the redshift of the maximum energy injection, 
similar to the case of the monomial potential.  The remaining
parameter, which we take to be $f$, is undetermined. In particular it
is not clear which value of $f$ corresponds to the best fit fluid
model. For figure~\ref{fig:comparecos} we fix $f = 0.05 M_{\rm pl}$. 

Fixing the height and redshift of energy injection relates the initial
field value with $f$. 
For $f \simeq M_{\rm pl}$ the initial field value required to get the best fit
injection shape is much smaller than $f$, so that the cosine potential
reduces to the monomial potential studied in the previous section.
As $f$ is reduced, the required initial field approaches $\pi f$, so
that small values of $f$ require a fine-tuned initial condition.
Furthermore, for  $\phi_i \gtrsim f$, $V''(\phi_i) < 0$ corresponding
to an instability in the spectrum of fluctuations.
For $f \lesssim 0.01 M_{\rm pl}$, the instability causes the
fluctuations to grow so rapidly that linear perturbation theory
becomes invalid.
For our choice of $f = 0.05M_{\rm Pl}$ the fluctuations remain under
control, but the initial field value must be
tuned at the $\mathcal{O}(1\%)$ level. 

In figure~\ref{fig:comparecos} we show two choices for the cosine potential. 
The ``fluid match $\cos$'' model (shown in dashed red) is chosen to match the
background energy injection shape of the best fit fluid model as
closely as possible. The remaining cosmological parameters were taken
from the best fit values in~\cite{Poulin:2018cxd}. We also present the ``best fit
$\cos$'' model (shown in solid red) in which all parameters are
freely floated, but does not correspond to the best fit fluid model in
any way.
In the fluid case we use the public
code developed by~\cite{poulin_code} to implement the fluctuations. For the
$\phi^4$ and cosine potentials we use the equations of motion of the
scalar field to derive the cosmological evolution of fluctuations.

We show our results for background cosmology in each of the cases
described above in figure~\ref{fig:comparecos}. We see that at the background level the
``fluid match $\cos$'' model does follow the best fit fluid model at late
times. However, it is signifcantly different at the peak of the energy
injection. The ``best fit $\cos$'' model has a much lower peak
injection energy than all the other models, reflecting the relative inability
of the CMB to
accommodate large energy injections of this form. In particular, the
energy injection in the ``best fit $\cos$'' model does not resemble
that of the fluid model at all.
We plot the residuals of the temperature
power spectrum with respect to a reference $\Lambda$CDM model. The models
that were fit to the CMB, i.e. $\phi^4$, ``best fit $\cos$'' and the
``best fit fluid'' model all have small residuals (as expected from the fact
that these models were chosen by minimizing CMB likelihoods). However,
the ``fluid match $\cos$'' model has enormous residuals in a range of
$\ell$ corresponding to modes which enter the horizon around the peak
of the energy injection.  This is quite consistent with our argument above: at
the onset of oscillations, the fluid model is in no way an
approximation to the $\cos$ potential.

Finally, an interesting question is if there is \emph{any} model that
does reproduce the fluid model.  Even if one did
reproduce the smooth energy injection of the fluid model, it is not
clear at all that the fluctuations in that particular model have
the same sound speed as the fluid model.  In particular, we can
(numerically) derive a potential for a scalar field such that the
energy injection is identical to the fluid model.
However, this potential has a region for which $V''(\phi) < 0$, and
so the fluctuations in this region are unstable.
It would seem that it is an open question whether the
fluid model can ever be realized in a realistic physics model.

\bibliography{ref.bib}
\bibliographystyle{utphys}

\end{document}